\pdfoutput=1 % pour la compilation pdflatex par arXiv
\documentclass[smallextended,numbook]{svjour3}

\journalname{}

\usepackage{ulem}

\usepackage{graphicx,float,amsmath,amssymb,mathrsfs}

\usepackage[english,francais]{babel}

\usepackage{url,hyperref}  
\usepackage{cite}

\usepackage[usenames,dvipsnames]{xcolor}

\definecolor{Cblue}{HTML}{045FB4}
\definecolor{Cred}{HTML}{DF0101}

\hypersetup{
  colorlinks=true, %set true if you want colored links
  linktoc=true,     %set to all if you want both sections and
  linkcolor=Cred,
  citecolor=Cblue, 
  urlcolor=Cblue
}

\renewcommand{\em}{\it}

\definecolor{M_Beige}         {rgb}{0.96 , 0.96 , 0.86}

\definecolor{M_Brown}         {rgb}{0.65 , 0.16 , 0.16}

\definecolor{M_Gold}          {rgb}{1.00 , 0.84 , 0.00}

\definecolor{M_LemonChiffon}  {rgb}{1.00 , 0.98 , 0.80}

\definecolor{M_Orange}        {rgb}{1.00 , 0.60 , 0.00}

\definecolor{M_Pink}          {rgb}{0.80 , 0.55 , 0.60}

\definecolor{M_Violet}          {rgb}{0.83 , 0.21 , 0.93}

\definecolor{M_Green}          {rgb}{0.2 , 0.6 , 0.2}

\definecolor{M_Gray}          {rgb}{0.7 , 0.7 , 0.7}

\definecolor{M_BluPal}          {rgb}{0.7 , 0.7 , 0.9}

\renewcommand{\leq}{\leqslant}
\renewcommand{\geq}{\geqslant}

\newcommand{\ket}[1]{|\kern.3ex#1\kern.3ex\rangle}
\newcommand{\bra}[1]{\langle\kern.3ex #1 \kern.3ex|}
\newcommand{\scalar}[2]{\langle\kern.3ex{#1}\kern.3ex|\kern.3ex{#2}\kern.3ex\rangle}
\newcommand{\mean}[1]{\left\langle #1\right\rangle}

\newcommand{\EXP}[1]{\mathrm{e}^{#1}}         % exponentielle

\def\dd{{\rm d}}                  % la differenciation
\def\D{{\mathcal{D}}}                 % D integrale fonctionnelle

\def\Xint#1{\mathchoice
  {\XXint\displaystyle\textstyle{#1}}%
  {\XXint\textstyle\scriptstyle{#1}}%
  {\XXint\scriptstyle\scriptscriptstyle{#1}}%
  {\XXint\scriptscriptstyle\scriptscriptstyle{#1}}%
  \!\int}
\def\XXint#1#2#3{{\setbox0=\hbox{$#1{#2#3}{\int}$}
    \vcenter{\hbox{$#2#3$}}\kern-.5\wd0}}

\def\dashint{\Xint-}

\newcommand{\abs}[1]{\left| #1 \right|}

\newcommand{\diagram}[3]{\raisebox{#3}{\includegraphics[scale=#2]{#1}}}

\newcommand\antiddots{\mathinner{\mkern2mu\raise1pt\hbox{.}\mkern2mu
    \newline \raise4pt\hbox{.}\mkern2mu\raise7pt\hbox{.}\mkern1mu}}

\def\Ni{N_1}

\def\lagNl{\mu_0^{(1)}}
\def\lagNr{\mu_0^{(2)}}
\def\lagS{\mu_1}

\def\Pn{P_{N,\kappa}}

\def\rhoS{\rho^\star}
\def\rhoLag{\rhoS}
\def\rhoz{\rho_0^\star}

\def\sz{s_0}

\def\C{c}
\def\CM{G}

\def\rhoL{\rho_1}
\def\rhoR{\rho_2}

\begin{document}

\selectlanguage{english}

%%%%%%%%%%%%%%%%%%%%%%%%%%%%%%%%%%%%%%%%%%%%%%%%%%%%%%%%%%%%%%%%%%%%%%%%%%%%% 
\renewcommand{\labelitemi}{$\bullet$}
\renewcommand{\labelitemii}{$\star$}
%%%%%%%%%%%%%%%%%%%%%%%%%%%%%%%%%%%%%%%%%%%%%%%%%%%%%%%%%%%%%%%%%%%%%%%%%%%%% 

\title{Truncated linear statistics associated with the top eigenvalues of random matrices}

\author{Aur\'elien Grabsch \and Satya N. Majumdar \and Christophe Texier}

\institute{
  Aur\'elien Grabsch \and Satya N. Majumdar \and Christophe Texier
  \at
  LPTMS, CNRS, Univ. Paris-Sud, Universit\'e Paris-Saclay, 91405 Orsay, France\\
}

\date{May 16, 2018}

\maketitle

\begin{abstract}
  Given a certain invariant random matrix ensemble characterised by
  the joint probability distribution of eigenvalues
  $P(\lambda_1,\ldots,\lambda_N)$, many important questions have been
  related to the study of linear statistics of eigenvalues
  $L=\sum_{i=1}^Nf(\lambda_i)$, where $f(\lambda)$ is a known
  function.  We study here truncated linear statistics where the sum
  is restricted to the $N_1<N$ largest eigenvalues:
  $\tilde{L}=\sum_{i=1}^{N_1}f(\lambda_i)$.  Motivated by the analysis
  of the statistical physics of fluctuating one-dimensional
  interfaces, we consider the case of the Laguerre ensemble of random
  matrices with $f(\lambda)=\sqrt{\lambda}$.  Using the Coulomb gas
  technique, we study the $N\to\infty$ limit with $N_1/N$ fixed. We
  show that the constraint that
  $\tilde{L}=\sum_{i=1}^{N_1}f(\lambda_i)$ is fixed drives an infinite
  order phase transition in the underlying Coulomb gas. This
  transition corresponds to a change in the density of the gas, from a
  density defined on two disjoint intervals to a single interval.  In
  this latter case the density presents a logarithmic divergence
  inside the bulk.  Assuming that $f(\lambda)$ is monotonous, we show
  that these features arise for any random matrix ensemble and
  truncated linear statitics, which makes the scenario described here
  robust and universal.

\end{abstract}

% \pacs{05.60.Gg ; 03.65.Nk ; 05.45.Mt ; 72.15.Rn}

% \pacs{73.20.Fz}{Weak or Anderson localisation}

% 02.50.-r Probability theory, stochastic processes, and statistics
% 02.50.Cw 	Probability theory 
% 02.50.Ey 	Stochastic processes 
% 03.65.Nk    Scattering theory 
% 05.10.Gg 	Stochastic analysis methods (Fokker-Planck, Langevin, etc.) 
% 05.40.-a 	Fluctuation phenomena, random processes, noise, and
% 05.40.Jc Brownian motion
% 05.45.Mt    Quantum chaos ; semiclassical methods 
% 05.60.Gg    Quantum transport
% 05.60.-k 	Transport processes
% 05.70.Np  Interface and surface thermodynamics
% 72.   Electronic transport in condensed matter
% 72.10.-d   Theory of electronic transport; scattering mechanisms
% 72.10.Bg   General formulation of transport theory 
% 72.15.Rn Localization effects (Anderson or weak localization)

% 73.   Electronic structure and electrical properties of surfaces, interfaces,
% thin films, and low-dimensional structures 
% 73.23.-b     Electronic transport in mesoscopic systems 
% 73.20.Fz Weak or Anderson localization

\vspace{0.25cm}


%%%%%%%%%%%%%%%%%%%%%%%%%%%%%%%%%%%%%%%%%%%%%%%%%%%%%%%%%%%%%%%%%%%%%%%%%%%%%%%%%%%%%%%%%% 
%%%%%%%%%%%%%%%%%%%%%%%%%%%%%%%%%%%%%%%%%%%%%%%%%%%%%%%%%%%%%%%%%%%%%%%%%%%%%%%%%%%%%%%%%% 

\section{Introduction}
\label{sec:Introduction}

Introduced in physics by Wigner and Dyson in the 1950s in order to
model the complexity in atomic nucleus, random matrix theory has
irrigated many fields of physics, ranging from electronic quantum
transport
\cite{Bee97,GuhMulWei98,AleBroGla02,MelKum04,Bro95,MelBar99,SomWieSav07,VivMajBoh08,KhoSavSom09,VivMajBoh10,VV08,GraTex15,CFVJointStatEPL},
quantum information (entanglement in random bipartite quantum states)
\cite{Page93,FMPPS08,PFPPS10,FFPPY13,NadMaj10,NadMajVer11} or
statistical physics of fluctuating interfaces \cite{NadMaj09,Nad11}
(see \cite{MajSch14} for a review).  The first questions arising in
nuclear physics were related to the statistical analysis of the
spectrum (level distribution and correlations), encoded in the joint
probability density function of the eigenvalues $P(\lambda_1, \ldots,
\lambda_N)$ characterizing the matrix ensemble.  Another class of
questions has arisen later, concerning the statistical properties of
linear statistics of the eigenvalues
\begin{equation}
  L = \sum_{n=1}^N f(\lambda_n)
  \:,
  \label{eq:LS0}
\end{equation}
where $f$ is a given function (not necessarily linear).  Many physical
quantities can be expressed under such a form, as illustrated in the
aforementioned references.  Several tools have been developed to
tackle this problem within \textit{invariant} random matrix
ensembles,\footnote{ The case where the matrix distribution is
  invariant under changes of basis, i.e. when eigenvalues and
  eigenvectors are uncorrelated.  }  such as orthogonal polynomials,
Selberg's integrals or the Coulomb gas method. Although orthogonal
polynomials method can be used to compute the moments of $L$ and
yields explicit formulae for the characteristic function in terms of
determinants, these results are quite difficult to use in practice, in
particular to study the limit $N \to \infty$.  In this limit, the
Coulomb gas technique proves to be the most efficient method for the
analysis of the full distribution (in particular the large deviation
tails characterizing the atypical fluctuations).  The idea is to
interpret the eigenvalue distribution $P(\lambda_1, \ldots,
\lambda_N)$ as the Gibbs measure for a one-dimensional (1D) gas of
particles, ``charges'', with logarithmic interactions (eigenvalues
then correspond to the positions of the particles) \cite{DysonI}.
% \cite{DysonI,DysonII,DysonIII}.
For references in the mathematical literature,
cf.~\cite{AroGui97,AroZei98}.  The analysis of the distribution of the
linear statistics $L$ is then mapped onto the determination of the
configuration of charges that minimizes the energy of the gas under
the constraint that $L = \sum_{n} f(\lambda_n)$ is fixed.  In the
thermodynamic limit, $N \to \infty$, the density of eigenvalues can be
considered as continuous, which makes the optimization problem
solvable by several techniques such as by resolvent
method~\cite{BreItzParZub78} or using the Tricomi theorem
\cite{DeaMaj06}.  This Tricomi's theorem was found very useful and was
first used to obtain the large deviation function associated to the
distribution of the largest eigenvalue, say $\lambda_1$, of Gaussian
matrices \cite{DeaMaj06,DeaMaj08} and Wishart matrices \cite{VMB07}.
This problem can be related to the study of a linear statistics of the
form \eqref{eq:LS0} as the cumulative distribution of the largest
eigenvalue coincides with the probability that all the eigenvalues are
below the threshold, thus $\mathrm{Proba}\{\lambda_1\leq
W\}=\mathrm{Proba}\{L=N\}$ for $f(\lambda) = \Theta(W-\lambda)$, where
$\Theta$ is the Heaviside step-function.  The problem has been further
generalised to consider the number of eigenvalues in an arbitrary
interval, $f(\lambda)=\mathbf{1}_{[a,b]}(\lambda)$
\cite{MajNadScaViv09,MajNadScaViv11,MarMajSchViv14}, a question
relevant in various contexts, like principal component analysis in
statistics \cite{MajViv12} or particle-number fluctuations of fermions
in a harmonic trap at zero temperature~\cite{MarMajSchViv14a,MMSV16}.

An interesting aspect of the statistical analysis of linear statistics
is the possibility of phase transitions in the Coulomb gas, driven by
the constraint, which correspond to transitions in the density of the
optimal charge configuration: for example the splitting of the density
or the transition between a soft edge and a hard edge (density
vanishing or diverging at a boundary).  Several examples were studied
in
Refs.~\cite{VivMajBoh08,NadMaj09,VivMajBoh10,TexMaj13,NadMaj10,NadMajVer11,MajSch14,GraTex15,CFVJointStatEPL}
(see also Table~\ref{tab:CGpt} in the conclusion).

In this paper, we consider a new kind of constraint arising from the study of linear statistics restricted to the $\Ni < N$ \textit{largest} eigenvalues:
\begin{equation}
  \tilde{L} = \sum_{n=1}^{\Ni} f(\lambda_n)
  \hspace{1cm}
  \mbox{with }
  \lambda_1 > \lambda_2 > \cdots > \lambda_N
  \:.
  \label{eq:defTrLS}
\end{equation}
This new problem interpolates between the two important types of
questions mentioned above:
\begin{itemize}
\item for $\Ni = 1$ and $f(\lambda)=\lambda$, the distribution of
  $\tilde{L}$ corresponds with the distribution of the maximal
  eigenvalue $\lambda_1$.
\item The case $\Ni = N$ reduces to the statistical analysis of the full linear statistics~$L$. 
\end{itemize}
We introduce the fraction $\kappa=\Ni/N$ of charges contributing to
the truncated linear statistics and rescale this latter as
$s=N^{-\eta}\,\tilde{L}$, where the exponent $\eta$ controls the large
$N$ scaling of $\tilde{L}$ (i.e. $\tilde{L}\sim N^{\eta}$)~; the
precise value of $\eta$ depends on the ensemble and the function $f$,
as we will see later.  The main question is here to determine the
distribution
\begin{align}
  \label{eq:defPnsInteg}
  &P_{N,\kappa}(s) 
  =N!
  \\\nonumber
  &\times
  \int\dd\lambda_1\int^{\lambda_1}\dd\lambda_2\cdots\int^{\lambda_{N-1}}\dd\lambda_N\,
  P(\lambda_1,\ldots,\lambda_N)\:
  \delta\!\left(
    s - N^{-\eta} \sum_{n=1}^{\Ni} f(\lambda_n)
  \right)
\end{align}
(we omit to specify the precise domain of integration, which depends on the matrix ensemble).

Although our results are very general and can be applied to
many linear statistics or matrix ensembles, we will focus on an
example involving Wishart matrices. 
Such matrix ensembles were
introduced by Wishart \cite{Wis28} for the study of empirical
covariance in multivariate statistics. They correspond to matrices
of the form $Y = X^\dagger X$, with $X$ of size $M \times N$ with
independent and identically distributed Gaussian entries.  It is
convenient to introduce the Dyson index $\beta$ corresponding to
real ($\beta=1$), complex ($\beta=2$) or quaternionic ($\beta=4$)
matrix entries. The eigenvalues of $Y$ are positive, and have the
following joint probability density function:
\begin{equation} P(\lambda_1, \ldots, \lambda_N) \propto \prod_{i<j}
  \abs{\lambda_j - \lambda_j}^\beta \prod_{n=1}^N
  \lambda_n^{\frac{\beta}{2}(M-N+1)-1} \EXP{-\beta \lambda_n/2},
  \hspace{1cm} \lambda_n > 0 \:.
  \label{eq:jpdfWish}
\end{equation} In this expression, $M\geq N$ is an integer, however
we will consider below an extension where $M-N+1$ is replaced by a
real positive number (Laguerre ensemble of random matrix theory).  In
Ref.~\cite{NadMaj09}, this distribution was shown to arise in the
analysis of a model of 1D interfaces \cite{Fisher84,Einstein2003},
which is described in Section~\ref{sec:Model}.

\subsection{Main results}

We have analysed the distribution \eqref{eq:defPnsInteg} in the large $N$ limit with $\kappa=\Ni/N$ fixed for the Laguerre ensemble, Eq.~(\ref{eq:jpdfWish}) with $M-N=\mathcal{O}(1)$, and $f(\lambda) = \sqrt{\lambda}$.
The result can be written under the form~\footnote{ In the paper, the expressions for the limiting
  behaviours of $\Pn(s)$ must be understood more rigorously as
  $\lim_{N\to\infty}\big[-2/(\beta N^2)\big]\ln[\Pn(s)]=\Phi_\kappa(s)$.  } 
\begin{equation}
  \Pn(s) \underset{N\to\infty}{\sim} \exp \left\lbrace
    - \frac{\beta N^2}{2}  \Phi_\kappa(s)
  \right\rbrace
  \:,
\end{equation}
where the large deviation function has the following limiting behaviours:
\begin{equation}
  \Phi_\kappa(s) \simeq
  \
  \begin{cases}
    -2 \ln s
    & \text{as } s\to 0
    \\[0.15cm]
    \displaystyle
    \frac{\pi^2}{4 - c_0 + c_0 \ln c_0/4}(s-\sz(\kappa))^2
    & \text{as } s\to s_0(\kappa)
    \\[0.15cm]
    \displaystyle
    \frac{s^2}{\kappa} + \kappa(3\kappa-4) \ln s
    & \text{as } s\to +\infty
  \end{cases}
\end{equation}
$s_0(\kappa)$, given parametrically in Eqs.~(\ref{eq:optsC},\ref{eq:optkC}), denotes a critical line in the $(\kappa,s)$ plane, as shown in the phase diagram in Fig.~\ref{fig:PhDiag}.
The constant $c_0$ is controlled by $\kappa$, cf. Eq.~\eqref{eq:optkC} below.
A sketch of the distribution is plotted in Fig.~\ref{fig:SchemaDistr}.

\begin{figure}[!ht]
  \centering
  \includegraphics[width=0.6\textwidth]{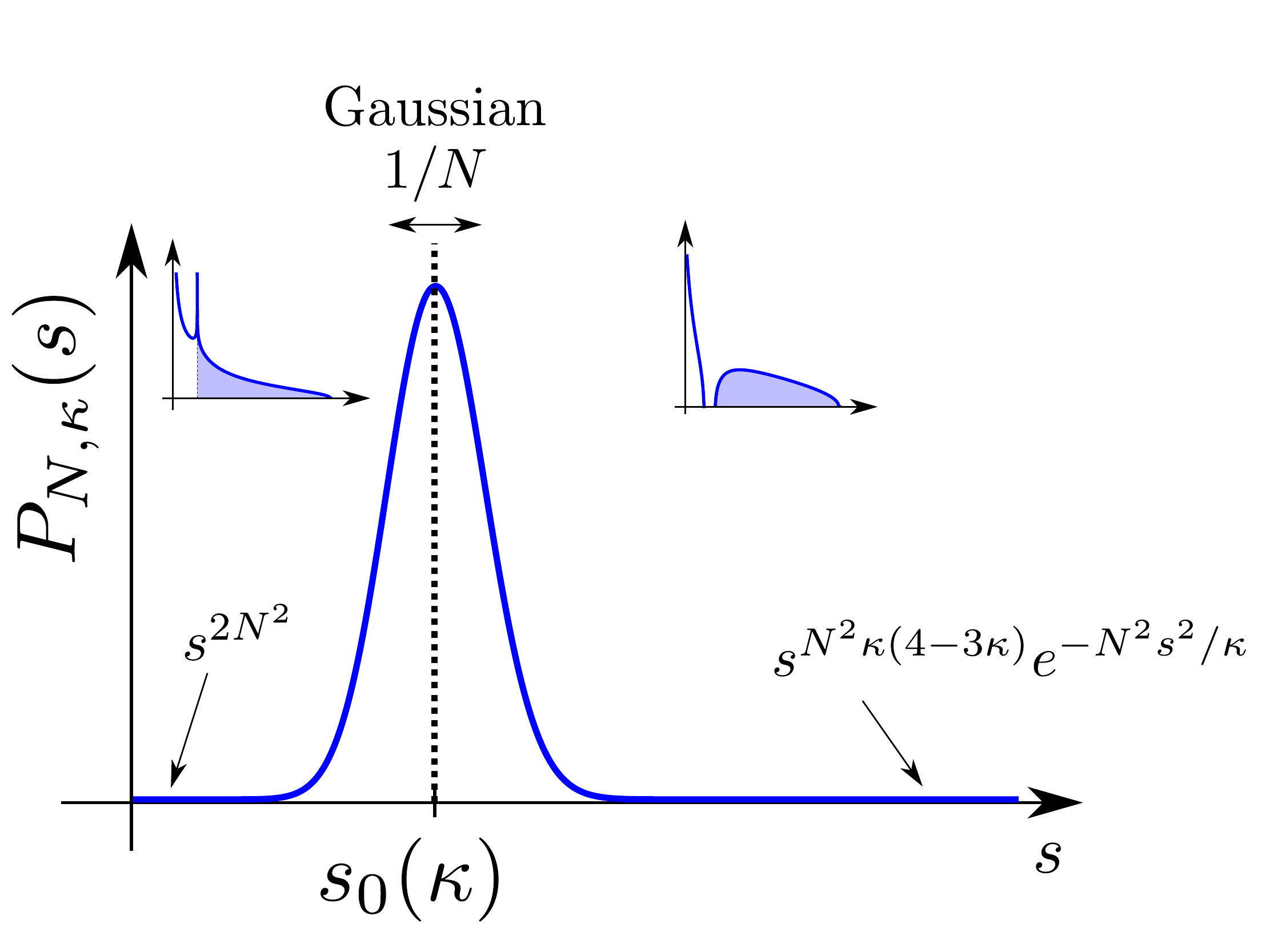}
  \caption{\it Sketch of the distribution $\Pn(s)$ of the truncated linear statistics.}
  \label{fig:SchemaDistr}
\end{figure}

This specific distribution can be understood from a new  universal scenario that we now describe. 
In the $N\to\infty$ limit, the multiple
integrals~(\ref{eq:defPnsInteg}) are dominated by the optimal
configuration of charges of the Coulomb gas. The two important
parameters that govern the behaviour of the gas are the fraction
$\kappa=\Ni/N$ and the parameter $s$ controlling the constraint
$s=N^{-\eta}\,\sum_{n=1}^{\Ni}f(\lambda_n)$.  For a given $\kappa$,
the change of $s$ drives a phase transition in the Coulomb gas
corresponding to a change of the optimal charge density.  Phase 1
corresponds to a density supported on two disjoint intervals (see
Fig.~\ref{fig:PhDiag}).  As $s$ approaches the critical value
$\sz(\kappa)$, the gap between the two intervals shrinks. Exactly at
the transition, the density is a smooth function (see
Fig.~\ref{fig:PhDiag}).  If $s$ deviates from $\sz(\kappa)$,
entering in Phase 2, a logarithmic divergence emerges at the point
where the two intervals have merged (see Fig.~\ref{fig:PhDiag}).
The energy of the Coulomb gas exhibits an essential singularity at
the transition $s=s_0(\kappa)$, hence the phase transition is of \textit{infinite
  order}.

% The infinite order phase transition corresponds to an essential singularity in the large deviation function at $s=s_0(\kappa)$. 
% These results can be straightforwardly generalised to the cases $\beta=1$ or $4$. 

\begin{figure}[!ht]
  \centering
  \includegraphics[width=0.8\textwidth]{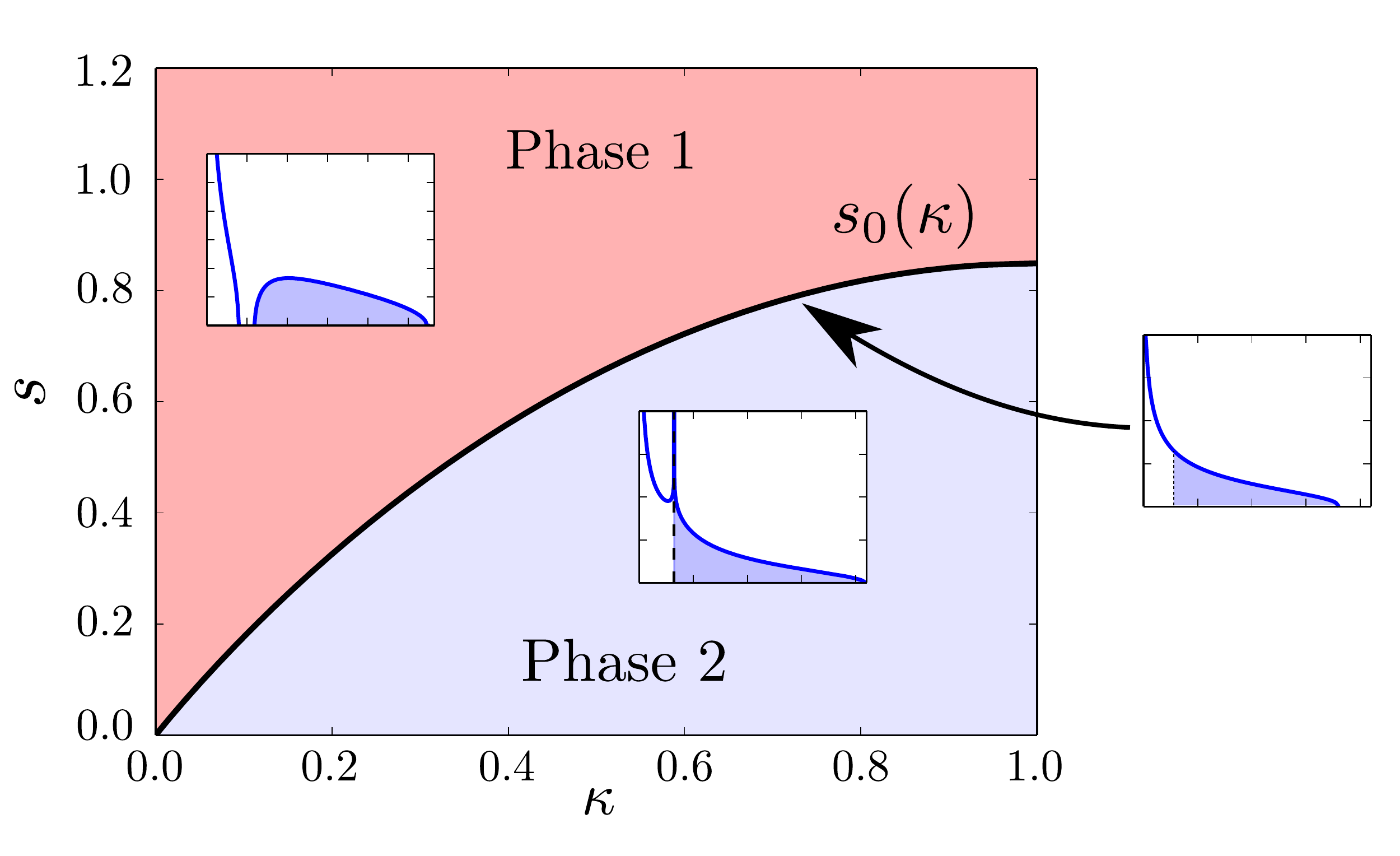}
  \caption{\it 
    Phase diagram for the Coulomb gas in the $(\kappa,s)$ plane, 
    where $\kappa=\Ni/N$ and $s=N^{-\eta}\sum_{n=1}^{\Ni} f(\lambda_n)$ 
    (for the case $f(\lambda)=\sqrt{\lambda}$ in the Laguerre ensemble, with $\eta=3/2$). 
    Insets show the shape of the corresponding optimal ``charge'' density profiles (shaded areas correspond to the $\Ni$ largest eigenvalues).
    The phase transition occurs on the line $\sz(\kappa)$ (thick black line) corresponding to the most probable value of $\sum_{n=1}^{\Ni} f(\lambda_n)$.}
  \label{fig:PhDiag}
\end{figure}

The scenario just described is not restricted to the particular case considered here (Laguerre ensemble with the function $f(\lambda)=\sqrt{\lambda}$). 
We have shown (Section~\ref{sec:Conclusion} and Appendix~\ref{app:sol}) that this scenario is \textit{universal} and holds for any monotonous function $f$ and any matrix ensemble, and can be summarised as follows.
For fixed $\kappa$, the optimal density of charges undergoes an infinite order phase transition at the critical value $s_0(\kappa)$:
\begin{itemize}
\item 
  In the first phase, the density is supported on two disjoint intervals.
\item 
  In the second phase, the density exhibits a logarithmic singularity. 
\end{itemize}
The general expressions of these densities can be found in Appendix~\ref{app:sol}.
Note that depending on the function $f$, other phase transitions might be present (cf. Section~\ref{sec:Conclusion}).

\subsection{Outline of the paper}

The paper is organised as follows: the interface model is described in Section~\ref{sec:Model}.  Section~\ref{sec:CoulombGas} presents the Coulomb gas analysis and the determination of the large deviation function. 
The implications of these results for the interface model are discussed in Section~\ref{sec:DistrS}. 
Subsection~\ref{subsec:numerics} describes the outcome of numerical simulations performed to check our analytical results.
The paper is closed with some concluding remarks and a brief discussion of another truncated linear statistics within the Jacobi ensemble.  
Appendix~\ref{app:sol} describes some technical issues and show that the main scenario is robust and not restricted to the specific truncated linear statistics studied in Sections~\ref{sec:CoulombGas} and~\ref{sec:DistrS}.

%%%%%%%%%%%%%%%%%%%%%%%%%%%%%%%%%%%%%%%%%%%%%%%%%%%%%%%%%%%%%%%%%%%%%%%%%%%%%%%%%%%%%%%%%% 
%%%%%%%%%%%%%%%%%%%%%%%%%%%%%%%%%%%%%%%%%%%%%%%%%%%%%%%%%%%%%%%%%%%%%%%%%%%%%%%%%%%%%%%%%% 

\section{The interface model}
\label{sec:Model}

Our discussion will be based on a model of $N$ non intersecting (1+1)-dimensional elastic fluctuating interfaces. This model was first introduced by de Gennes \cite{DeGen68}, and later by Fisher \cite{Fisher84} in order to study wetting of surfaces or the commensurate/incommensurate transition in deposition of atoms on a surface (for a review, cf.~\cite{Nad11}). 
The interfaces evolve on a substrate of size $L$, which induces a repulsive force. More precisely, each interface is described by its height $h_n(x)$, $n = 1, \ldots, N$, for $x$ between $0$ and $L$ and we assume periodic boundary conditions $h_n(L) = h_n(0)$. An energy can be associated to the system of $N$ interfaces:
\begin{equation}
  E[\lbrace h_n(x) \rbrace] = \sum_{n=1}^N \mathcal{E}[h_n(x)]
\end{equation}
where
\begin{equation}
  \mathcal{E}[h_n(x)] = \int_0^L \left[
    \frac{1}{2} \left( \frac{\dd h}{\dd x} \right)^2 + V(h(x))
  \right] \dd x
  \:.
\end{equation}
$(1/2)\int \left({\dd h_n}/{\dd x} \right)^2$ is the elastic energy and $V(h_n)$ some external potential. 
Following the approach of Ref.~\cite{NadMaj09} we choose it in the form
\begin{equation}
  V(h) = \frac{b^2 h^2}{2} + \frac{\alpha(\alpha-1)}{2h^2}
  \hspace{1cm}
  \text{with } b>0 \text{ and } \alpha > 1
  \:.
\end{equation}
\begin{figure}[!ht]
  \centering
  \includegraphics[width=0.5\textwidth]{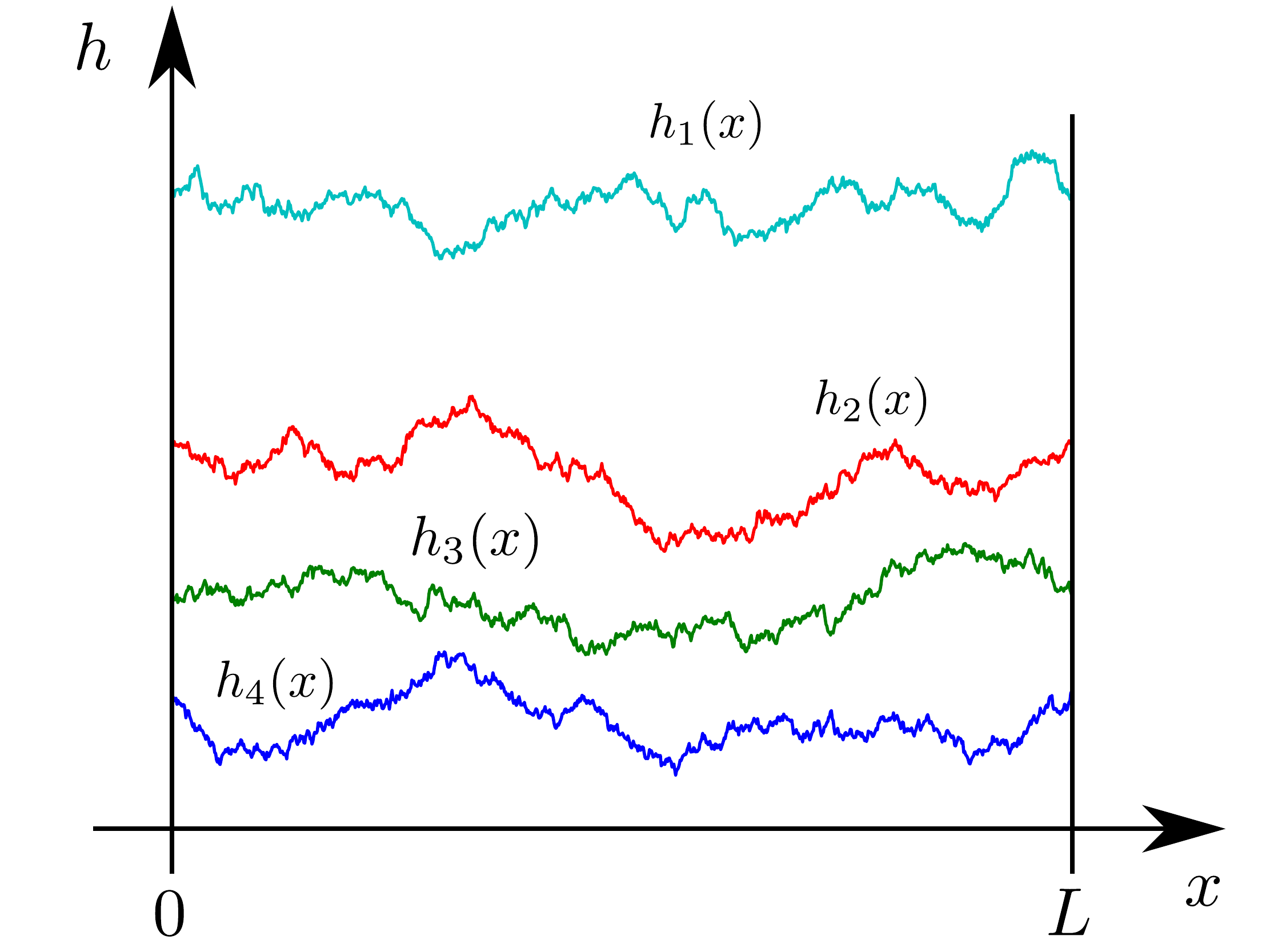}
  \caption{\it Nonintersecting Brownian interfaces $h_n(x)$ with periodic boundary conditions $h_n(L) = h_n(0)$.}
  \label{fig:Interf}
\end{figure}
This potential is made of two terms: a harmonic confining term and a repulsion term from the substrate (supposed
at $h=0$), which ensures that the interfaces remain in the region $h > 0$. This repulsion in $h^{-2}$ is justified by entropic considerations \cite{Fisher84,NadMaj09}.
The model with purely harmonic confinement was used in order to study vicinal surfaces of cristals (for a review, see Ref.~\cite{Einstein2003}).
At thermal equilibrium, a configuration of $N$ interfaces $\lbrace h_n(x) \rbrace_{n=1}^N$ can be associated to a Boltzmann weight:
\begin{equation}
  \mathcal{P}[\lbrace h_n(x) \rbrace] \propto
  \exp \left\lbrace
    - \frac{1}{k_\mathrm{B} T} \, E[\lbrace h_n(x) \rbrace]
  \right\rbrace
  \:,
\end{equation}
where $T$ is the temperature and $k_\mathrm{B}$ the Boltzmann constant. We set $k_\mathrm{B} T = 1$ for convenience. Moreover we impose that the interfaces do not intersect, hence we can order them: $h_1 > h_2 > \cdots > h_N$. Our choice of boundary conditions implying translational invariance, the joint distribution of the heights is independent of the position $x$. This distribution was obtained in Ref.~\cite{NadMaj09} by mapping the heights of the interfaces to the positions of free quantum particles trapped in the potential $V(h)$. Since the interfaces cannot intersect, the fictitious particles are fermions. 
In the limit $L \to \infty$, their distribution is controlled by the wave function $\Psi_0(h_1,\ldots,h_N)$ of the many body ground state, which yields \cite{NadMaj09}:
\begin{equation}
  P_\mathrm{interfaces}(h_1, \ldots, h_N) \propto 
  \prod_{i<j} (h_i^2 - h_j^2)^2 \prod_{n=1}^N h_n^{2\alpha}\, \EXP{- b h_n^2}
  \:.
  \label{eq:jpdfInterfaces}
\end{equation}
The Vandermonde determinant $\prod_{i<j}(h_i^2 - h_j^2)^2$ originates in the fermionic nature of the particles, equivalently, in the condition that the interfaces do not cross. 
A simple change of variable  $\lambda_n = b\, h_n^2$ allows to relate this distribution to the joint probability density function of eigenvalues for $N \times N$ Wishart matrices (Laguerre ensemble):
\begin{equation}
  P(\lambda_1, \ldots, \lambda_N) \propto
  \prod_{i<j} (\lambda_i - \lambda_j)^2 \prod_{n=1}^N \lambda_n^{\alpha-1/2}\, \EXP{- \lambda_n}
  \:.
  \label{eq:jpdfWishart}
\end{equation}
It corresponds to Eq.~\eqref{eq:jpdfWish} with $\beta=2$ and $M = N + \alpha - 1/2$. 
This particular value of $\beta=2$ is not related to the breaking of the time reversal symmetry but can be here understood from the relation between the wave function, given by a Slater determinant, and the probability density 
$P_\mathrm{interfaces}(h_1,\ldots,h_N)=|\Psi_0(h_1,\ldots,h_N)|^2$. 
The distribution \eqref{eq:jpdfWishart} will be the starting point of our analysis. 
We are interested in the distribution of the center of mass of the $\Ni$ highest interfaces:
\begin{equation}
  \CM = \frac{1}{\Ni} \sum_{n=1}^{\Ni} h_n
  \hspace{0.5cm} , \hspace{0.5cm}
  h_1 > h_2 > \cdots > h_N
  \:.
\end{equation}
Denote $\kappa = \Ni/N$ the fraction of interfaces we consider. This can be rewritten as
\begin{equation}
  \CM = \frac{1}{\Ni \sqrt{b}} \sum_{n=1}^{\Ni} \sqrt{\lambda_n}
  \hspace{0.5cm} , \hspace{0.5cm}
  \lambda_1 > \lambda_2 > \cdots > \lambda_N
  \:,
  \label{eq:Glambdas}
\end{equation}
i.e. the function in \eqref{eq:defTrLS} is 
\begin{equation}
  \label{eq:ChoiceF}
  f(\lambda) = \sqrt{\lambda}
\end{equation}
We will study the limit $N \to \infty$, with $0 < \kappa < 1$ fixed, using a Coulomb gas method.

%%%%%%%%%%%%%%%%%%%%%%%%%%%%%%%%%%%%%%%%%%%%%%%%%%%%%%%%%%%%%%%%%%%%%%%%%%%%%%%%%%%%%%%%%% 
%%%%%%%%%%%%%%%%%%%%%%%%%%%%%%%%%%%%%%%%%%%%%%%%%%%%%%%%%%%%%%%%%%%%%%%%%%%%%%%%%%%%%%%%%% 

\section{Coulomb gas analysis of the truncated linear statistics}
\label{sec:CoulombGas}

The Coulomb gas method consists in rewriting the joint distribution (\ref{eq:jpdfWishart}) as a Gibbs measure $\exp[- E_\mathrm{gas}]$, with the energy $E_\mathrm{gas} = - \sum_{ i \neq j} \ln \abs{\lambda_i - \lambda_j} + \sum_i (\lambda_i - (\alpha-1/2) \ln \lambda_i)$. This energy describes a gas of particles on a semi-infinite line, trapped in a confining potential $V(\lambda) = \lambda - (\alpha - 1/2) \ln \lambda$ and submitted to repulsive logarithmic interaction between each other. 
In the limit $N \to \infty$, the interaction energy scales with $N$ as $\sum_{i\neq j} \ln \abs{\lambda_i - \lambda_j} \sim N^2$, and the confinement energy as $\sum_i \lambda_i \sim N \lambda$, where $\lambda$ is a typical value taken by the $\lambda_i$'s. Since we expect the distribution of charges to find an equilibrium between confinement and repulsion, the eigenvalues should scale as $\lambda_i \sim N$. Hence, we rescale them as
\begin{equation}
  \lambda_i = N x_i
  \:.
\end{equation}
We can then introduce the empirical density
\begin{equation}
  \rho(x) = \frac{1}{N} \sum_{n=1}^N \delta(x -x_n)
  \:,
\end{equation}
and rewrite the measure (\ref{eq:jpdfWishart}) as a functional of this density (we neglect entropic contributions which are of order $N$, compared to the energy of order $N^2$ \cite{Dys62,DeaMaj08}):
\begin{equation}
  P(\lambda_1, \ldots, \lambda_N) \, \dd \lambda_1 \cdots \dd \lambda_N
  \rightarrow
  \EXP{- N^2  \mathscr{E}[\rho]} \, \D \rho
  \:,
\end{equation}
with the energy
\begin{equation}
  \mathscr{E}[\rho] = - \int \rho(x) \rho(y) \ln \abs{x-y} \, \dd x \dd y
  + \int \rho(x) \left( x - \frac{\alpha-1/2}{N} \ln x \right) \dd x 
  \:.
  \label{eq:energy}
\end{equation}
Since we consider only the large $N$ limit, we will drop the $1/N$ term in the energy, 
assuming that $\alpha=\mathcal{O}(N^0)$.
The rescaled linear statistics \eqref{eq:Glambdas} reads
\begin{equation}
  s = \frac{\kappa \sqrt{b} \, \CM}{\sqrt{N}} = \int_\C \rho(x) \sqrt{x} \, \dd x
  \:,
  \label{eq:defS}
\end{equation}
where $\C$ is a lower bound ensuring that only the $\Ni$ largest eigenvalues contribute to the integral, i.e.
\begin{equation}
  \int_\C \rho(x) \, \dd x = \kappa  
  \: .
  \label{eq:defCk}
\end{equation}
Our aim is to compute the distribution of the rescaled center of mass $s$, which can be expressed in terms of path integrals over the density:
\begin{align}
  \label{eq:PathIntegral}
  &	\Pn(s) 
  =
  \\\nonumber
  &\frac{\displaystyle \int \hspace{-0.1cm}\dd \C \hspace{-0.1cm}\int\hspace{-0.1cm} \mathcal{D}\rho\,
    \EXP{-N^2 \mathscr{E}[\rho]}\:
    \delta\! \left(\int_\C \hspace{-0.1cm}\dd x\, \rho(x)  - \kappa\right)
    \delta\! \left(\int^\C \hspace{-0.25cm}\dd x\,\rho(x) - (1-\kappa) \right)
    \delta\!\left(\int_\C\hspace{-0.1cm}\dd x\, \sqrt{x} \rho(x) - s\right)}
  {\displaystyle \int\hspace{-0.1cm}\dd x\, \dd \C \int\mathcal{D}\rho\,\EXP{-N^2 \mathscr{E}[\rho]}\:
    \delta\! \left(\int_\C\hspace{-0.1cm}\dd x\, \rho(x)  - \kappa\right)
    \delta\! \left(\int^\C\hspace{-0.25cm}\dd x\, \rho(x)  - (1-\kappa) \right)}
  % \:.
\end{align}
A dual problem was studied in Ref.~\cite{MajViv12}, where the distribution of the number of eigenvalues above a fixed threshold was considered. This corresponds to study the distribution of $\kappa$ with fixed $c$ corresponding to the threshold, while releasing the constraint on $s$. However, here we are interested in the distribution of $s$ with $\kappa$ fixed, and $c$ is now a parameter determined by $\kappa$. The addition of the constraint on $s$ will lead to a completely different phenomenology.

%%%%%%%%%%%%%%%%%%%%%%%%%%%%%%%%%%%%%%%%%%%%%%%%%%%%%%%%%%%%%%%%%%%%%%%%%%%%%%%%%%%%%%%%%% 

\subsection{Saddle point equations and large deviation function}

\begin{figure}[!ht]
  \centering
  \includegraphics[width=0.35\textwidth]{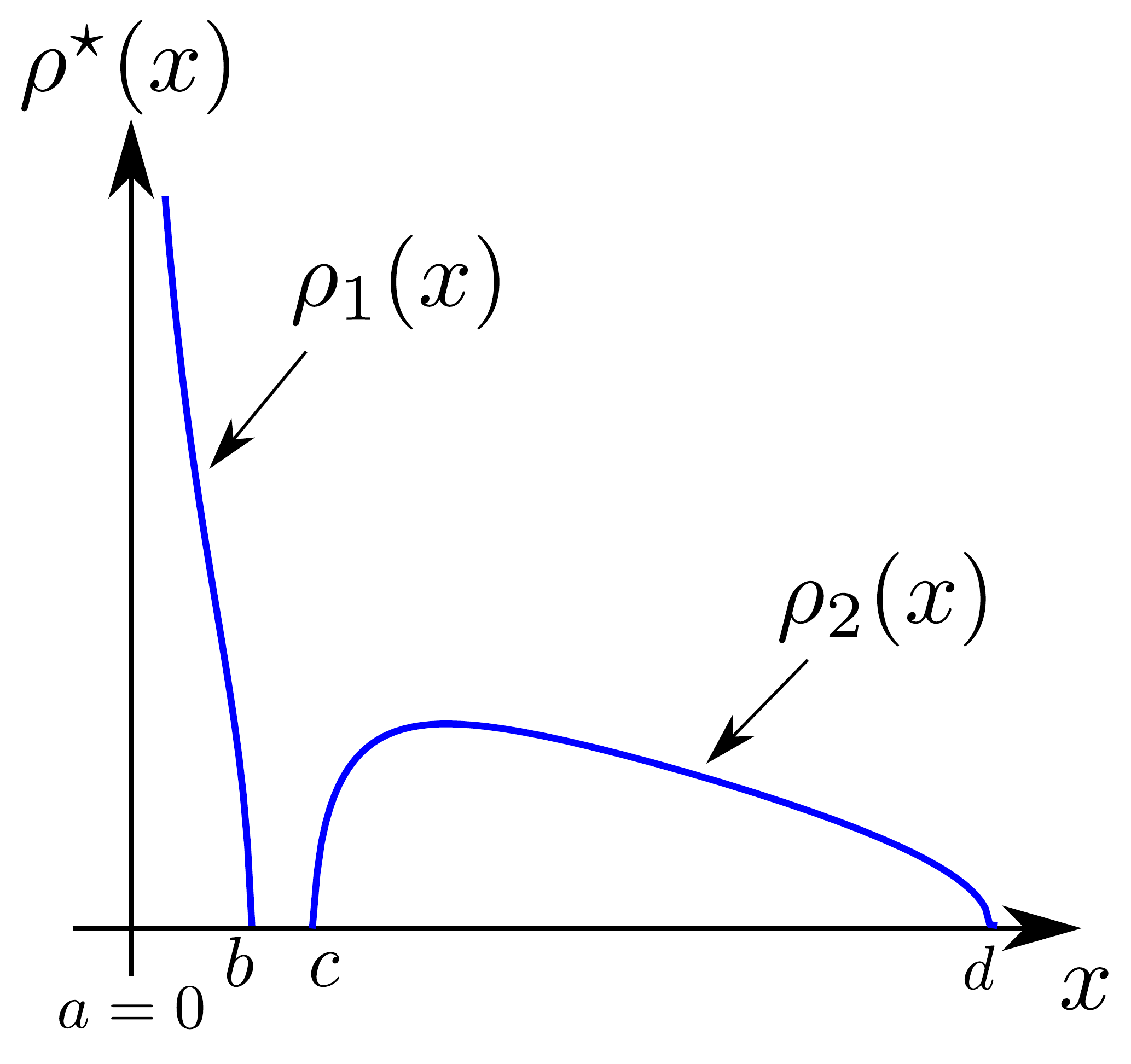}
  \caption{\it Densities $\rho_1$ and $\rho_2$ defined by Eq.~(\ref{eq:DefDensities}) in Phase~1. 
    % The support of $\rho_1$ is denoted $[a,b]$, the support of $\rho_2$ is $[c,d]$.
  }
  \label{fig:Dens}
\end{figure}

When $N \to \infty$, the path integrals are dominated by the minimum of the energy under the constraints imposed by the Dirac $\delta$-functions. These constraints can be handled by introducing Lagrange multipliers $\lagNl$, $\lagNr$ and $\lagS$. Denote the ``free energy''
\begin{align}
  \mathscr{F}[\rho;\lagNl, \lagNr, \lagS] 
  &= \mathscr{E}[\rho]
  + \lagNl \left(\int^\C \rho(x) \dd x - (1-\kappa) \right)
  \nonumber\\
  &+ \lagNr \left(\int_\C \rho(x) \dd x - \kappa \right)
  +\lagS \left(\int_\C \sqrt{x} \rho(x) \dd x - s\right)
  \:.
  \label{eq:FreeEnergy}
\end{align}
The numerator of Eq.~(\ref{eq:PathIntegral}) is dominated by the density of charges $\rhoS(x;\kappa,s)$ that minimizes $\mathscr{F}$. This density is solution of the saddle point equation
$$
\left.\frac{\delta \mathscr{F}}{\delta \rho(x)}\right|_{\rhoS} \!\!\! = 0 
$$
implying
\begin{equation}
  2 \int \rhoS(y;\kappa,s) \ln \abs{x-y} \dd y = x + 
  \left\lbrace
    \begin{array}{ll}
      \lagNl & \text{for } x < \C
      \\[0.125cm]
      \lagNr + \lagS \sqrt{x} & \text{for } x > \C
    \end{array}
  \right.
  \label{eq:SteepestDescent}
\end{equation}
which can be understood as the energy balance for the charge at point $x$ 
between the confinement and the logarithmic repulsion.
The Lagrange multipliers $\lagNl$ and $\lagNr$ correspond to chemical potentials fixing the fraction of eigenvalues respectively below and above $\C$. 
The term with $\lagS$ adds another external potential coming from the constraint on $s$. In order to solve this equation, we first take its derivative and interpret the resulting relation as the equilibrium of the forces exerted on the charge at position $x$:
\begin{equation}
  2 \dashint \frac{\rhoS(y;\kappa,s)}{x-y} \dd y = 1 + 
  \left\lbrace
    \begin{array}{ll}
      0 & \text{for } x < \C\\
      \frac{\lagS}{2 \sqrt{x}} & \text{for } x > \C
    \end{array}
  \right.
  \label{eq:SteepestDescentD}
\end{equation}
where the integral is a Cauchy principal value integral.
It is convenient to describe the density $\rhoS$ with two functions: $\rhoR$ related to the fraction $\kappa$ of eigenvalues under consideration, and $\rhoL$ for the others (see Figure \ref{fig:Dens}) % and \ref{fig:rhoZero}). Explicitly:
\begin{equation}
  \rhoR(x) = \frac{1}{N} \sum_{n=1}^{\Ni} \delta(x-x_n)
  \:,
  \qquad
  \rhoL(x) = \frac{1}{N} \sum_{n=\Ni+1}^{N} \delta(x-x_n)
  \:.
  \label{eq:DefDensities}
\end{equation}
The confining potential ensures that the eigenvalues remain in a bounded region in space, hence the densities $\rhoL$ and $\rhoR$ will have compact supports. Denote $[a,b]$ the support of $\rhoL$, and $[c,d]$ the support of $\rhoR$, where $c$ is the boundary introduced above in Eqs.~(\ref{eq:defS},\ref{eq:defCk}), as shown in Figure \ref{fig:Dens}. It is possible to have $b=c$, as shown in Fig.~\ref{fig:rho} (left part) and Fig.~\ref{fig:SchemaCcl}.

We rewrite Eq.~(\ref{eq:SteepestDescentD}) as
\begin{align}
  \label{eq:SDRho1}
  2 \int_a^b \frac{\rhoL(y)}{x-y} \dd y + 2 \dashint_c^d \frac{\rhoR(y)}{x-y} \dd y
  &=  1 + \frac{\lagS}{2 \sqrt{x}} 
  \hspace{1cm} 
  \text{for } x > \C 
  \\
  \label{eq:SDRho2}
  2 \dashint_a^b \frac{\rhoL(y)}{x-y} \dd y + 2 \int_c^d \frac{\rhoR(y)}{x-y} \dd y 
  &= 1 
  \hspace{2.25cm}  
  \text{for } x < \C
  \:.
\end{align}
Note that, in these two equations, the principal value is only needed when $x$ belongs to the support of the density in the integral.
These equations can be solved by a double iteration of a theorem due to Tricomi, as in Ref.~\cite{MajNadScaViv11}. This theorem provides an explicit inversion of Cauchy singular equations of the form
\begin{equation}
  \dashint \frac{\rho(y)}{x-y} \dd y = g(x)
  \:,
  \label{eq:Tri1}
\end{equation}
under the assumption that the solution has one single support $[a,b]$. This formula reads~\cite{Tri57}
\begin{equation}
  \rho(x) = \frac{1}{\pi \sqrt{(x-a)(b-x)}} \left\lbrace
    A + \dashint_a^b \frac{\dd t}{\pi} \frac{\sqrt{(t-a)(b-t)}}{t-x} g(t)
  \right\rbrace
  \:,
  \label{eq:Tri2}
\end{equation}
where $A$ is a constant. This procedure, detailed in Appendix~\ref{app:sol}, is quite cumbersome, but it is the only one available since the standard method using resolvent technique \cite{BreItzParZub78} is even more complicated in this case. 
The solution of these equations hence give $\rhoLag = \rhoL \cup \rhoR $. The parameters $\C$ and $\lagS$ are fixed by the constraints:
\begin{equation}
  \int_c^d \rhoR(x) \, \dd x = \kappa
  \hspace{1cm} , \hspace{1cm}
  \int_c^d \rhoR(x) \sqrt{x} \, \dd x = s
  \:.
  \label{eq:Constraints}
\end{equation}
These two equations give $\lagS$ and $\C$ as functions of $s$ and $\kappa$. Let us denote $\lagS^\star(\kappa,s)$ and $\C^\star(\kappa,s)$ the two solutions. The numerator of Eq.~\eqref{eq:PathIntegral} is dominated by the density $\rhoS$.
The optimal density for the denominator can be obtained straightforwardly by releasing the constraint on $s$, which can be done by setting $\lagS^\star(\kappa,s) = 0$. Solving this equation for $s$ gives $s = \sz(\kappa)$, and we will denote the corresponding density $\rhoz(x)=\rhoS(x;\kappa,\sz(\kappa))$.
Finally, Eq.~\eqref{eq:PathIntegral} yields
\begin{equation}
  \Pn(s) \underset{N \to \infty}{\sim}
  \exp \left\lbrace
    - N^2 \Phi_\kappa(s)
  \right\rbrace
  \:,
\end{equation}
where we have introduced the large deviation function
\begin{equation}
  \Phi_\kappa(s) = \mathscr{E}[\rhoS(x;\kappa,s)] - \mathscr{E}[\rho_0^\star(x)]
  \:.
\end{equation}
This is the difference of energy between the two optimal configurations of charges dominating the numerator and the denominator of \eqref{eq:PathIntegral}, respectively. 
These energies are given by Eq.~\eqref{eq:energy}. A direct computation of the double integral is quite difficult in practice. However, an important simplification based on a ``thermodynamic'' identity was introduced in Ref.~\cite{GraTex15} and discussed in details in Refs.~\cite{CunFacViv16,GraTex16}:
\begin{equation}
  \frac{\dd \mathscr{E}[\rhoS(x;\kappa,s)]}{\dd s} = - \lagS^\star(\kappa,s)
  \:.
  \label{eq:thermoId}
\end{equation}
Using this relation one can obtain the large deviation function directly by integration of the Lagrange multiplier $\lagS$:
\begin{equation}
  \label{eq:thermoId2}
  \Phi_\kappa(s) = \int_s^{\sz} \lagS^\star(\kappa,t) \, \dd t
  \:.
\end{equation}
We will make extensive use of this relation in our study of the distribution $\Pn(s)$. 
To lighten the notations, the dependences on the parameters $\kappa$ and $s$ will be implicit from now on. In particular the optimal density will be simply denoted by $\rhoS(x)$.

%%%%%%%%%%%%%%%%%%%%%%%%%%%%%%%%%%%%%%%%%%%%%%%%%%%%%%%%%%%%%%%%%%%%%%%%%%%%%%%%%%%%%%%%%% 

\subsection{Optimal density without constraint}

In the absence of the constraint ($\lagS = 0$), the steepest descent equation (\ref{eq:SteepestDescentD}) reduces to
\begin{equation}
  2 \dashint \frac{\rhoz(y)}{x-y} \dd y = 1
  \:,
\end{equation}
which can be solved straightforwardly using Tricomi's theorem (\ref{eq:Tri1},\ref{eq:Tri2}). The density is given by the Mar\v{c}enko-Pastur distribution \cite{MP67}:
\begin{equation}
  \rhoz(x) = \frac{1}{2 \pi} \sqrt{\frac{4-x}{x}}
  \:.
\end{equation}
The divergence of this density at $x=0$ corresponds to an accumulation of charges, or equivalently of interfaces, near the origin. It has a compact support, meaning that the largest value will typically be $x_1\sim4$. Using the relation $\lambda_1 = N x_1 = b h_1^2 $, we deduce that the highest interface will fluctuate around the position 
\begin{equation}
  \label{eq:HighestInterface}
  h_1 \sim 2 \sqrt{N/b}
\end{equation}

It is clear from Eq.~\eqref{eq:thermoId2} that this distribution is associated to the maximum of the probability, hence to the most probable value of $s$. This value is given by
\begin{equation}
  \sz(\kappa) = \int_{c_0}^4 \rhoz(x) \sqrt{x} \, \dd x = \frac{(4-c_0)^{3/2}}{3\pi}
  \:,
  \label{eq:optsC}
\end{equation}
where $c_0$ is fixed by the fraction of charges under consideration:
\begin{equation}
  \kappa = \int_{c_0}^4 \rhoz(x) \, \dd x =
  \frac{2}{\pi} \arccos \frac{\sqrt{c_0}}{2} - \frac{\sqrt{c_0(4-c_0)}}{2\pi}
  \:.
  \label{eq:optkC}
\end{equation}
Solving the second equation for $c_0$ and plugging the result into the first gives the most probable value $\sz(\kappa)$ taken by the truncated linear statistics, or equivalently, by the center of mass of the highest interfaces. It is the solid line represented in Figure \ref{fig:PhDiag}. In particular, one gets the following asymptotics:
\begin{align}
  \kappa & \simeq \frac{(4-c_0)^{3/2}}{6 \pi}
  &
  \text{ as } c_0 \to 4
  \:,
  \\
  \kappa & \simeq 1 - \frac{2 \sqrt{c_0}}{\pi}
  &
  \text{ as } c_0 \to 0
  \:,
\end{align}
from which we deduce:
\begin{align}
  \sz(\kappa) & \simeq 2 \kappa - \frac{3(3\pi)^{2/3}}{10\times2^{1/3}}\,\kappa^{5/3}  
  &
  \text{ as } \kappa \to 0
  \:,
  \\
  \sz(\kappa) & \simeq \frac{8}{3\pi} - \frac{\pi}{4}(1-\kappa)^2 
  &
  \text{ as } \kappa \to 1
  \:.
\end{align}
We will see later that $\sz(\kappa)$ defines a phase transition line (Fig.~\ref{fig:PhDiag}).

%%%%%%%%%%%%%%%%%%%%%%%%%%%%%%%%%%%%%%%%%%%%%%%%%%%%%%%%%%%%%%%%%%%%%%%%%%%%%%%%%%%%%%%%%% 

\subsection{Solution of the saddle point equation}

We follow the procedure used in Ref.~\cite{MajNadScaViv11} in a different context.
We first solve Eq.~\eqref{eq:SDRho2} for $\rhoL$ using Tricomi's theorem. This gives a solution in terms of $\rhoR$, which can be plugged into Eq.~\eqref{eq:SDRho1} to obtain an equation on $\rhoR$ only. Then, this equation can be solved using Tricomi's theorem again to get $\rhoR$. Finally $\rhoL$ can be deduced from the solution of Eq.~\eqref{eq:SDRho2}. This computation is carried out explicitly in Appendix~\ref{app:sol} in a more general case.

The solution takes a rather simple form, and can be expressed in a compact way in terms of $\rhoLag = \rhoL \cup \rhoR$. The additional term $\lagS \sqrt{x}$ in the potential coming from the constraint pushes the charges either towards or away from the origin depending on the sign of $\lagS$. 
Therefore we will distinguish these two cases, determined by the sign of $\lagS$.

\subsubsection{Phase 1: $\lagS < 0$ and  $s > \sz(\kappa)$ (cf. Fig.~\ref{fig:PhDiag})}

\begin{figure}[!ht]
  \centering
  \includegraphics[width=0.49\textwidth]{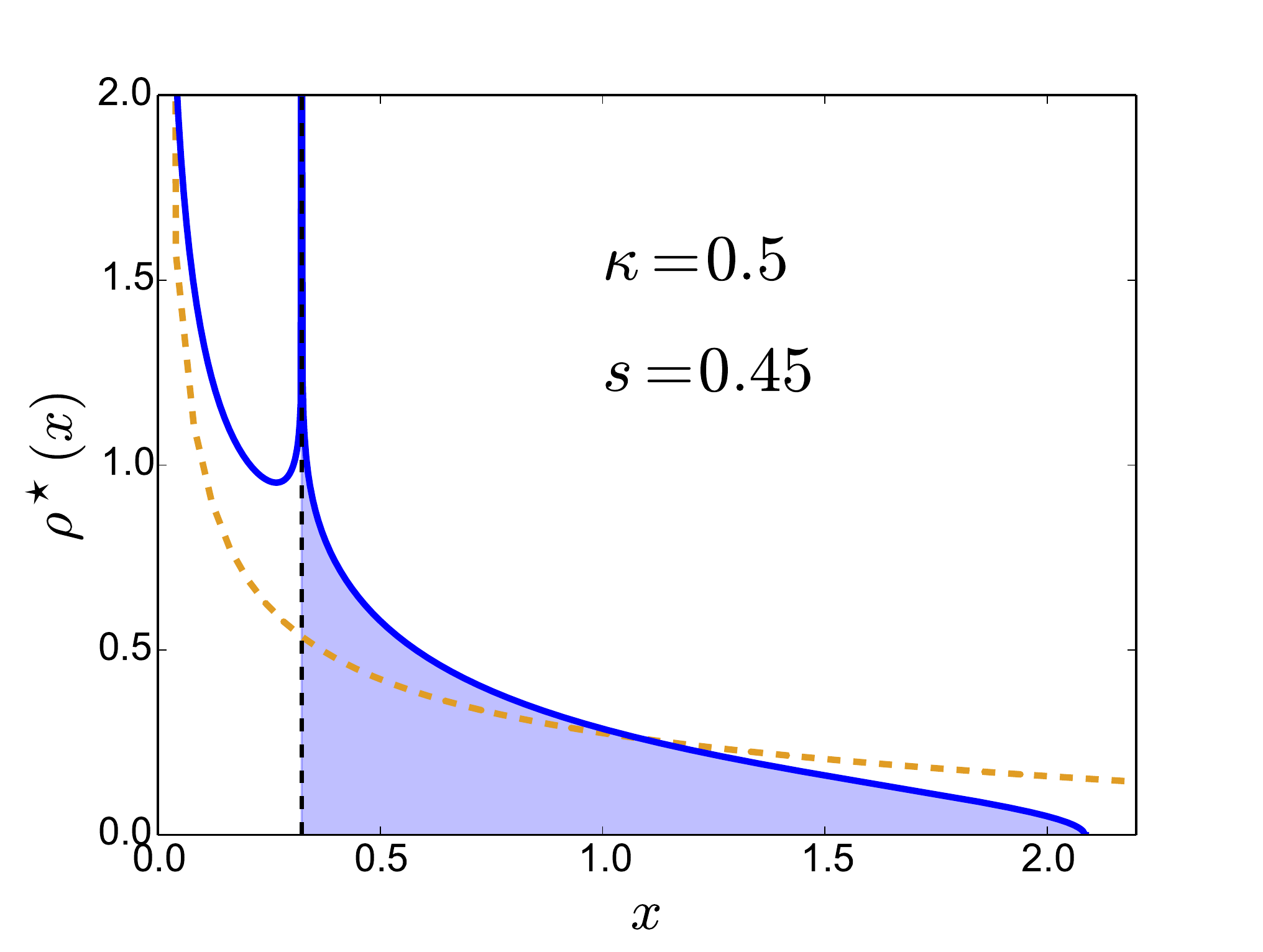}
  \includegraphics[width=0.49\textwidth]{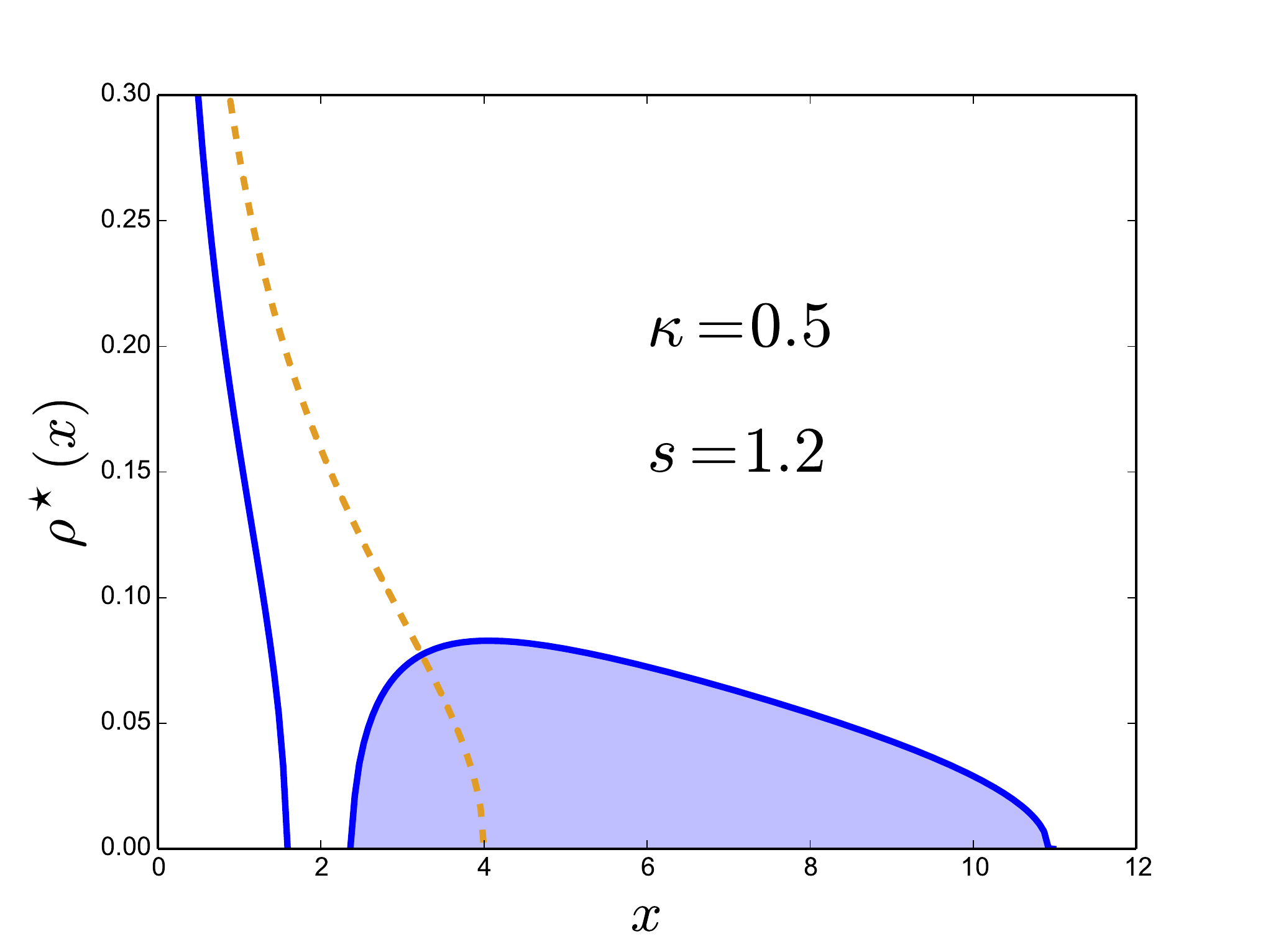}
  \caption{\it Optimal density $\rho^\star$ (plain line), compared to the density in the absence of constraint $\rho_0^\star$ (dashed). 
    On the left, $\lagS>0$, with $\kappa=0.5$ and $s=0.45$ (Phase~2). 
    On the right, the case $\lagS < 0$, with $\kappa=0.5$ and $s=1.2$ (Phase~1). 
    The dashed vertical line delimits the eigenvalues under consideration from the others. At this point, the distribution exhibits a logarithmic singularity.}
  \label{fig:rho}
\end{figure}

We have found the solution of Eqs.~(\ref{eq:SDRho1},\ref{eq:SDRho2}),
\begin{equation}
  \rhoLag(x) = \frac{\lagS}{2\pi^2} \frac{\mathrm{sign}(x-c)}{\sqrt{d-b}}
  \sqrt{\frac{(c-x)(b-x)}{x(d-x)}}
  \Pi \left( \frac{d-c}{d-x}, \sqrt{\frac{d-c}{d-b}} \right)
  \:,
  \label{eq:rhoPh1}
\end{equation}
which involves the complete elliptic integral of the third kind, defined by \cite{Grad}
\begin{equation}
  \Pi(n,k) = \int_0^1 \frac{\dd t}{(1-nt^2) \sqrt{(1-t^2)(1- k^2 t^2)}}
  \:,
\end{equation}
where the integral must be understood in a principal value sense if $n
> 1$.  The fact that the density has a simple analytical form relies
on the specific choice \eqref{eq:ChoiceF}, which is made clear by
inspection of (\ref{eq:Rho1Appendix},\ref{eq:Rho2Appendix})~: in this
case we have $f'(\lambda)=1/(2\sqrt{\lambda})$, which simplifies the
integral in (\ref{eq:Rho1Appendix},\ref{eq:Rho2Appendix}) when the
boundary is $a=0$.  Note that the same formula \eqref{eq:rhoPh1}
describes the density on the two disjoint intervals $[0,b]$ and
$[c,d]$ (see Fig.~\ref{fig:rho}, right). To obtain
Eq.~(\ref{eq:rhoPh1}), we have imposed that the general form of the
density obtained from Tricomi's theorem vanishes at $x=b$, $c$ and $d$
($\rhoLag(d)=0$ is ensured by the elliptic integral). These conditions
translate into the following equations:
% One should impose that the density $\rhoLag$ vanishes at $x=b$, $c$
% and $d$. However the equation coming from $\rhoLag(c)=0$ has already
% been used in the derivation of (\ref{eq:rhoPh1}).  There remain only
% the conditions from the vanishing at $x=b$,
\begin{equation}
  1 + \frac{c-b-d}{4} = \frac{\mu_1}{2\pi} \left[
    \sqrt{d-b} \, \mathrm{E} \left( \sqrt{\frac{d-c}{d-b}} \right)
    - \frac{c-b}{\sqrt{d-b}} \, \mathrm{K} \left( \sqrt{\frac{d-c}{d-b}} \right) 
  \right]
  \:,
  \label{eq:ph1a}
\end{equation}
coming from the vanishing at $x=b$, and
\begin{equation}
  1 + \frac{\mu_1}{\pi} \frac{1}{\sqrt{d-b}} \, \mathrm{K} \left( \sqrt{\frac{d-c}{d-b}} \right) = 0
  \:,
  \label{eq:ph1b}
\end{equation}
which arises from the vanishing at $x=d$ (the condition at $x=c$ is
already encoded in the previous two expressions). We have denoted
$\mathrm{K}$ and $\mathrm{E}$ the complete elliptic integrals of the
first and second kind respectively \cite{Grad}. The last free
parameters are determined by the two
constraints~(\ref{eq:Constraints}).
% 
% \begin{equation}
%   \int_c^d \rhoLag(x) \, \dd x = \kappa	\hspace{1cm} , \hspace{1cm}\int_c^d \rhoLag(x) \sqrt{x} \, \dd x = s
%   \label{eq:ph1Constr}
% \end{equation}
The energy balance of the Coulomb gas (\ref{eq:SteepestDescent}) gives some insight to understand why the density splits: the additional potential term coming from the constraint is $\lagS \sqrt{x}$. When $\lagS < 0$, this term tends to drive the charges away from the origin. But since this potential affects only the fraction $\kappa$ of the rightmost charges, only these charges are pulled to the right, while the others stay near the origin. Consequently, this gives a value of $s$ larger than the most probable one $\sz(\kappa)$. This is coherent with Eq.~\eqref{eq:thermoId}, which implies that the energy $\mathscr{E}[\rhoS]$ associated to this solution increases with $s$, hence a probability decaying as $s$ increases (with $s>\sz$).

\subsubsection{Phase 2: $\lagS > 0$ and $s < \sz(\kappa)$ (cf. Fig.~\ref{fig:PhDiag})}

Remarkably, the solution of the saddle point equation
(\ref{eq:SDRho1},\ref{eq:SDRho2}) takes a simple analytical form in
this case:
\begin{equation}
  \rhoLag(x) = \frac{1}{2 \pi} \sqrt{\frac{d-x}{x}} + \frac{\mu_1}{4\pi^2 \sqrt{x}} 
  \ln \frac{\sqrt{d-c} + \sqrt{d-x}}{\abs{\sqrt{d-c} - \sqrt{d-x}}}
  \:.
  \label{eq:rhoPh2}
\end{equation}
The density has one compact support $[0,d]$ (see Fig.~\ref{fig:rho},
left), indicating that the two densities $\rhoL$ and $\rhoR$ merge at
$b=c$. To obtain Eq.~(\ref{eq:rhoPh2}), we have imposed that the
general solution coming from Tricomi's theorem vanishes at $x=d$. This
results in the condition
% The parameters $c$, $d$ and $\lagS$ are fixed by $\rhoLag(d) = 0$,
\begin{equation}
  1 - \frac{d}{4} = \frac{\lagS}{2\pi} \sqrt{d-c}
  \:.
  \label{eq:ph2a}
\end{equation}
The remaining free parameters are fixed by the
constraints~(\ref{eq:Constraints}), which read explicitly:
\begin{align}
  \label{eq:ph2k}
  \kappa &= \frac{d}{2\pi} \arccos \sqrt{\frac{c}{d}} - \frac{\sqrt{c(d-c)}}{2\pi}
  + \frac{\lagS}{2\pi^2} \left( \sqrt{c} \, \ln \frac{c}{d} +
    2 \sqrt{d-c} \, \arccos \sqrt{\frac{c}{d}}
  \right)
  \:,
  \\
  \label{eq:ph2s}
  s &= \frac{(d-c)^{3/2}}{3\pi} + \frac{\lagS}{2\pi^2}(d-c)
  \:.
\end{align}
The density exhibits a logarithmic divergence at $x=c$ where $\rhoL$ and $\rhoR$ merge:
\begin{equation}
  \rhoLag(x) \underset{x \to c}{\simeq} -\frac{\lagS}{4\pi^2 \sqrt{c}} \ln \abs{x-c}
  \:.
\end{equation}
This singularity is clearly visible on Figure \ref{fig:rho} (left). It is unusual in the framework of the Coulomb gas to obtain such a divergence in the density of eigenvalues. A logarithmic behaviour has already been found at a hard edge in Ref.~\cite{NadMaj09}, where the density diverges as $-\ln x /\sqrt{x}$ at the origin. The distribution (\ref{eq:rhoPh2}) is, to the best of our knowledge, the first example presenting a purely logarithmic divergence \textit{in the bulk} of a density of eigenvalues. As before, we can interpret this solution in terms of the Coulomb gas: the additional potential $\lagS \sqrt{x}$ felt by the rightmost charges is pushing them towards the origin. But since they must remain to the right of the other charges, the whole density is pushed towards the origin. This phenomenon, caused by the new type of mixed constraints (\ref{eq:Constraints}), is at the origin of the singularity. 
Although we have discussed this mechanism on the example provided by our model of interfaces, it can be generalized easily to other linear statistics and matrix ensembles (cf. Section~\ref{sec:Conclusion} and Appendix~\ref{app:sol}).

\subsubsection{Infinite order phase transition}
\label{sec:PhTr}

From the point of view of the Coulomb gas, the parameter $s$ drives a phase transition from a phase where the density is supported on two disjoint intervals ($\lagS < 0$) to a phase supported by a single interval ($\lagS > 0$). 
The transition occurs when $\lagS = 0$, which corresponds to the line $s = \sz(\kappa)$ in the $(\kappa,s)$ plane, determined by (\ref{eq:optsC},\ref{eq:optkC}) (see Fig. \ref{fig:PhDiag}). On this line, the density is given by the Mar\v{c}enko-Pastur distribution. One can show that the large deviation function $\Phi_\kappa(s)$ and all its derivatives are continuous at this point (Appendix~\ref{app:sol}). 
However, $\Phi_\kappa(s)$ (and also $\mathscr{E}[\rhoS(x)]$) is non analytic.
Whereas the function admits a Taylor expansion on one side ($s=\sz^-$), there is an additional essential singularity on the other side of the transition ($s=\sz^+$):
\begin{equation}
  \label{eq:EssSingPhi}
  \Phi_\kappa(\sz + \epsilon) - \Phi_\kappa(\sz - \epsilon) = \mathcal{O}(\epsilon \, \EXP{-\gamma/\epsilon})
  \:,
\end{equation}
where
\begin{equation}
  \label{eq:ExprGamma}
  \gamma = \frac{\sqrt{4-c_0}}{\pi}(4-c_0 + c_0 \ln (c_0/4)) 
  \:,
\end{equation}
where $c_0$ is obtained from (\ref{eq:optkC}) (a derivation of this
result is given in Appendix~\ref{sec:AppPhTr}). Therefore, in the
standard terminology of statistical physics, this corresponds to a
phase transition of \textit{infinite order}.  Note that, due to the
\textit{specific} choice of the function $f(x)=\sqrt{x}$, a similar
singularity was already found in Ref.~\cite{NadMaj09}, where the
linear statistics (not truncated, i.e. for $\kappa=1$) is considered.
Let us emphasize that the existence of the essential singularity in
the large deviation function of truncated linear statistics
($\kappa<1$) is a much more \textit{universal} phenomenon, independent
of the choice of $f$, as demonstrated in Appendix~\ref{app:sol}.

%%%%%%%%%%%%%%%%%%%%%%%%%%%%%%%%%%%%%%%%%%%%%%%%%%%%%%%%%%%%%%%%%%%%%%%%%%%%%%%%%%%%%%%%%% 
%%%%%%%%%%%%%%%%%%%%%%%%%%%%%%%%%%%%%%%%%%%%%%%%%%%%%%%%%%%%%%%%%%%%%%%%%%%%%%%%%%%%%%%%%% 

\section{Distribution of the center of mass of the highest interfaces}
\label{sec:DistrS}

We first recall the main results obtained in Ref.~\cite{NadMaj09} concerning the distribution of the center of mass of the interfaces, which corresponds to set $\Ni=N$ (i.e. $\kappa=1$).
The mean value and the variance were found to be 
\begin{equation}
  \label{eq:NadalMajumdarPRE2009a}
  \mean{ \CM } \simeq \frac{8\sqrt{N}}{3\pi\sqrt{b}} 
  \hspace{0.5cm}\mbox{and} \hspace{0.5cm}
  \mathrm{Var}( \CM ) \simeq \frac{2}{\pi^2\,N\,b}
\end{equation}
and the large deviation tails 
\begin{equation}
  \label{eq:NadalMajumdarPRE2009b}
  P_{N,1}(s) \sim 
  \left\{
    \begin{array}{ll}
      s^{2N^2}      & \mbox{for } s\to0 \\
      \EXP{-N^2s^2} & \mbox{for } s\to\infty
    \end{array}
  \right.
\end{equation} 

The case $\kappa<1$ is studied by using the solutions $\rhoLag$ from the previous section. 
The large deviation function $\Phi_\kappa(s)$  can be easily obtained from the Lagrange multiplier $\lagS$ thanks to Eq.~\eqref{eq:thermoId2}.

%%%%%%%%%%%%%%%%%%%%%%%%%%%%%%%%%%%%%%%%%%%%%%%%%%%%%%%%%%%%%%%%%%%%%%%%%%%%%%%%%%%%%%%%%% 

\subsection{Typical values and variance}

The typical value taken by the truncated linear statistics $s$ is the one obtained by relaxing the constraint $\lagS=0$. 
It corresponds to the value $\sz(\kappa)$ defined by Eqs.~(\ref{eq:optsC},\ref{eq:optkC}). 
In the limit $N\to\infty$, typical and mean value coincide, thus the mean center of mass of the $\Ni$ highest interfaces is 
\begin{equation}
  \label{eq:MeanG}
  \mean{ \CM } = \frac{\sqrt{N}}{\kappa\sqrt{b}} \, \sz(\kappa)
  \simeq
  \frac{1}{\sqrt{b}}\times
  \left\{
    \begin{array}{ll}
      2\sqrt{N} 
      - \frac{3(3\pi)^{2/3}}{10\times2^{1/3}}\,\frac{\Ni^{2/3}}{N^{1/6}}  
      & \mbox{for } \kappa=\frac{\Ni}{N}\to0        
      \\[0.25cm]
      \frac{8\sqrt{N}}{3\pi} 
      \left[ 
        1+\frac{N-\Ni}{N} %+\mathcal{O}((1-\kappa)^2)
      \right] 
      &\mbox{for } \kappa \to 1^-
    \end{array}
  \right.
\end{equation}
In the limit $\kappa\to0$, the center of mass is close to the position of the highest interface,  Eq.~\eqref{eq:HighestInterface}.
In the limit $\kappa\to1$, the leading term corresponds to \eqref{eq:NadalMajumdarPRE2009a}, $\mean{ \CM }$ being slightly increased by removing the contribution of the lowest interfaces.
The variance can be obtained from a Taylor expansion of $\Phi_\kappa$ at $\sz(\kappa)$. 
As discussed in Section~\ref{sec:PhTr}, the large deviation function has an essential singularity at this point, hence its derivatives are the same on both sides and we can restrict the analysis to the case $\lagS > 0$ where the expressions are simpler.
% , and it will be valid on both sides. 
In the limit $\lagS \to 0^+$, Eqs.~(\ref{eq:ph2a},\ref{eq:ph2k},\ref{eq:ph2s}) yield
\begin{align}
  c &= c_0 + \frac{\lagS}{\pi \sqrt{4-c_0}} \ln \frac{c_0}{4} + \mathcal{O}(\lagS^2),\\
  d &= 4 - \frac{2\lagS}{\pi} \sqrt{4-c_0} + \mathcal{O}(\lagS^2),\\
  s &= \sz - \frac{\lagS}{2\pi^2} \left( 4 -c_0  + c_0 \ln
    \frac{c_0}{4} \right) + \mathcal{O}(\lagS^2)
  \label{eq:LinkSMuNearS0}
  \:,
\end{align}
The last expression gives $\lagS$ in terms of $s$, hence:
\begin{equation}
  \Phi_\kappa(s) = \frac{\pi^2}{4 - c_0 + c_0 \ln c_0/4}(s-\sz)^2 + \mathcal{O}((s-\sz)^3)
  \:.
\end{equation}
The distribution of $s$ exhibits a Gaussian peak around $\sz$,
\begin{equation}
  \Pn(s) \underset{s\sim \sz}{\sim} \exp \left\lbrace
    -\frac{N^2\pi^2}{4 - c_0 + c_0 \ln c_0/4}(s-\sz)^2
  \right\rbrace
  \:,
\end{equation}
from which we deduce the variance:
\begin{equation}
  \mathrm{Var}(s) = \frac{1}{2\pi^2N^2} \left( 4 -c_0  + c_0 \ln \frac{c_0}{4} \right)
  \:,
\end{equation}
where $c_0$ is determined by the value of $\kappa$ through Eq.~\eqref{eq:optkC}. 
The scaling of the fluctuations as $1/N$, i.e. much smaller than $1/\sqrt{N}$ for independent variables, is a signature of the long range correlations in the Coulomb gas.
Coming back to the interface model, we find the corresponding limiting behaviours for the variance of the center of mass: 
\begin{equation}
  \label{eq:VarG}
  \mathrm{Var}( \CM ) 
  \simeq      \frac{1}{b}\times
  \left\{
    \begin{array}{ll}
      \frac{6^{4/3}}{16\pi^{2/3}\,N^{1/3}\Ni^{2/3}}
      & \mbox{for } \kappa\to0  
      \\[0.25cm]
      \frac{2}{\pi^2\,N}  \left[1+\frac{N-\Ni}{N}\right]
      &\mbox{for } \kappa \to 1^-
    \end{array}
  \right.
\end{equation}
The leading term when $\kappa\to1$ corresponds to \eqref{eq:NadalMajumdarPRE2009a}.

It is quite interesting to compare the limiting behaviours obtained in the regime $\Ni\ll N$ ($\kappa\to0$) to the extreme statistics for the interfaces positions.
The position the uppermost interface $h_1$ was shown to be $h_1=2\sqrt{N/b} + \delta h_1$, where the typical fluctuations are described by the Tracy-Widom distribution with the scaling $\delta h_1\sim N^{-1/6}$~\cite{NadMaj09} (the large deviations of $h_1$ are also discussed in this paper).
The same scaling $N^{-1/6}$ that appears in the distribution of the maximum height distribution, also appears both in the subleading correction term in $\mean{\CM}$ and in the fluctuations.

%%%%%%%%%%%%%%%%%%%%%%%%%%%%%%%%%%%%%%%%%%%%%%%%%%%%%%%%%%%%%%%%%%%%%%%%%%%%%%%%%%%%%%%%%% 

\subsection{Large deviations}

\subsubsection{Limit $s \to \infty$} 

This corresponds to the case $\lagS \to -\infty$. In this limit, the fraction $\kappa$ of the rightmost charges are pushed to infinity, while the others remain close to the origin (Fig.~\ref{fig:rho}, right). Explicitly, it corresponds to $c,d \to \infty$, with $\frac{d-c}{d-b} \to 0$. Combining Eqs.~(\ref{eq:Constraints},\ref{eq:ph1a},\ref{eq:ph1b}) yields
\begin{align}
  b &= 4(1-\kappa) + 8 \kappa (25\kappa-4) \frac{1}{\lagS^2} + \mathcal{O}(\lagS^{-3})
  \:,
  \\
  c &= \frac{\lagS^2}{4} + \sqrt{2\kappa} \, \lagS + 4-\kappa + \mathcal{O}(\lagS^{-1})
  \:,
  \\
  d &= \frac{\lagS^2}{4} - \sqrt{2\kappa} \, \lagS + 4-\kappa + \mathcal{O}(\lagS^{-1})
  \:.
\end{align}
Rescaling $x = c + (d-c)y$, the density $\rhoR$ behaves as
\begin{equation}
  (d-c) \rhoR(x) = \frac{8\kappa}{\pi} \sqrt{y(1-y)}
  \left( 1 - \frac{3 \sqrt{2\kappa}}{\lagS} (2y-1) + \mathcal{O}(\lagS^{-2}) \right)
  \:.
  \label{eq:rhoAsympt}
\end{equation}
At leading order, the density $\rhoR$ is given by the semi-circle law. This is expected since the charges no longer feel the presence of the wall at the origin. Substituting $\rhoR(x)$ from Eq.~(\ref{eq:rhoAsympt}) in $s=\int \sqrt{x}\, \rhoR(x)\,\dd x$ gives
\begin{equation}
  \lagS = -\frac{2s}{\kappa} - \frac{\kappa(3\kappa-1)}{s} + \mathcal{O}(s^{-2})
  \:,
\end{equation}
hence after integration, Eq.~\eqref{eq:thermoId2},
\begin{equation}
  \Phi_\kappa(s) = \frac{s^2}{\kappa} + \kappa(3\kappa-4) \ln s + \mathcal{O}(1)
  \:,
\end{equation}
corresponding to the behaviour
\begin{equation}
  \Pn(s) \underset{s \to \infty}{\sim} s^{N^2 \kappa(4-3\kappa)} \EXP{-N^2 s^2/\kappa}
  \:.
\end{equation}
When $\kappa=1$, the dominant exponential term corresponds to \eqref{eq:NadalMajumdarPRE2009b}. 
In addition to the generalization to $\kappa < 1$, we also provide the next term, which gives the power law in the distribution of $s$. 
The exponential term can be obtained easily by a heuristic argument based on the Coulomb gas picture: for large $s$, the energy is dominated by the potential energy of the charges pushed to infinity. The typical value of the position of these charges is given by $s \sim \kappa \sqrt{x_\mathrm{typ}}$, corresponding to $x_\mathrm{typ} \sim s^2/\kappa^2$. The energy of these charges is estimated as $\mathscr{E}[\rhoS(x)] \sim \int_c \rhoS\, V \sim  \kappa \, V(x_\mathrm{typ}) \sim s^2/\kappa$.
{\sc Qed.}

\subsubsection{Limit $s \to 0$} 

In this case, all the charges are pushed towards the origin, meaning $0 < c < d \to 0$. In this limit, Eqs.~(\ref{eq:ph2a},\ref{eq:ph2s}) yield $\lagS \simeq 2\pi /\sqrt{d-c}$ and $s \simeq \lagS(d-c)/(2\pi^2)$. This gives the behaviour of the Lagrange multiplier $\lagS \simeq 2/s$. The large deviation function is again deduced by integration with respect to $s$, Eq.~\eqref{eq:thermoId2}:
\begin{equation}
  \Phi_\kappa(s) \simeq - 2 \ln s
  \hspace{1cm}
  \text{as } s \to 0
  \:.
\end{equation}
Equivalently, the distribution of $s$ behaves as
\begin{equation}
  \Pn(s) \underset{s \to 0}{\sim} s^{2N^2}
  \:.
\end{equation} 
Remarkably, the left tail of the distribution does not depend on the fraction $\kappa$ of eigenvalues considered, and coincides precisely with \eqref{eq:NadalMajumdarPRE2009b}. 
This can be understood as follows: in the limit $s \to 0$ all the charges are pushed towards the origin, so the energy of the Coulomb gas is dominated by the interaction term describing logarithmic repulsion among charges. The typical distance between two charges is $\delta x = x_\mathrm{typ}/N$, hence $\mathscr{E}[\rhoS(x)] \sim - \ln \delta x \sim -2 \ln s$.

%%%%%%%%%%%%%%%%%%%%%%%%%%%%%%%%%%%%%%%%%%%%%%%%%%%%%%%%%%%%%%%%%%%%%%%%%%%%%%%%%%%%%%%%%% 

\subsection{Numerics}
\label{subsec:numerics}

We have also performed Monte Carlo simulations of the Coulomb gas. 
One starts from a trial distribution of $N$ charges that fulfills the constraint \eqref{eq:defTrLS}.
Then, pairs of charges are randomly moved in a way that preserves the constraint \eqref{eq:defTrLS}, with acception or rejection rule according to the Metropolis algorithm. 
The density eventually relaxes towards the optimal density with minimum energy.
All the simulations were performed for $N=2000$ charges.

\begin{figure}[!ht]
  \centering
  \includegraphics[width=0.49\textwidth]{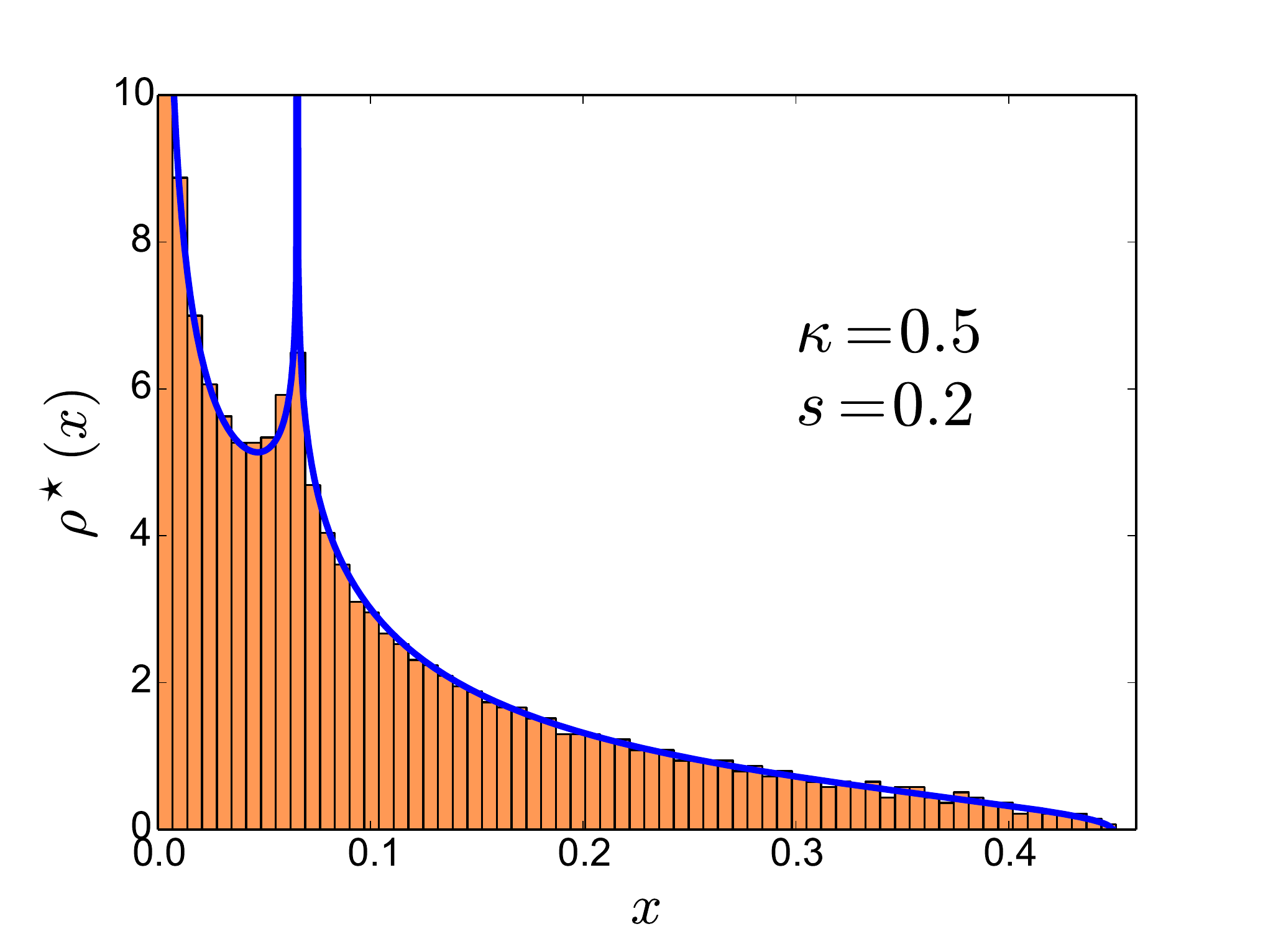}
  \includegraphics[width=0.49\textwidth]{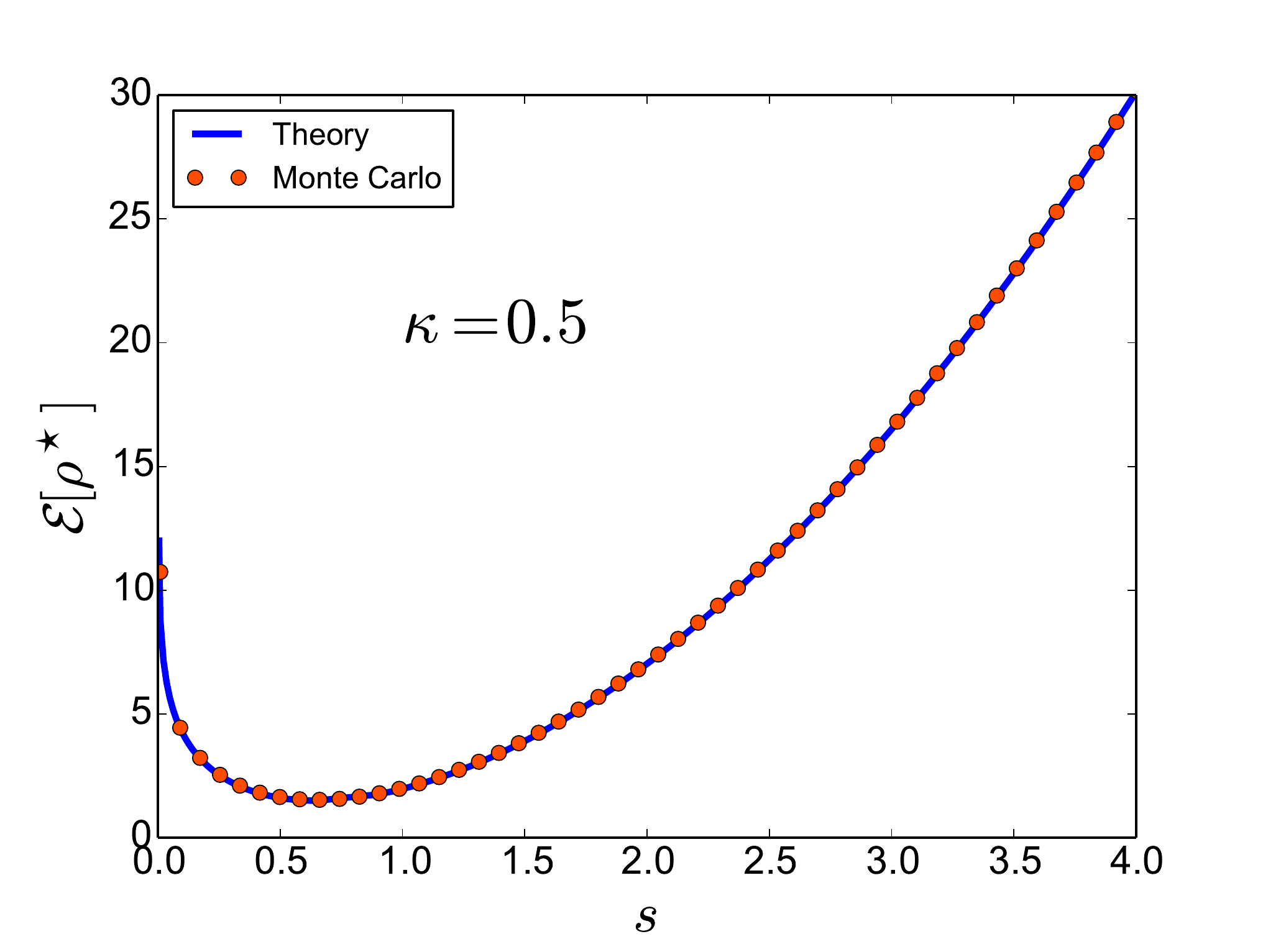}
  \caption{\it Monte Carlo simulations for the Coulomb gas, with $N=2000$ charges and $\kappa=0.5$. Left: histogram of the density obtained numerically for $s=0.2$ compared to the plot of Eq.~(\ref{eq:rhoPh2}) (no fit). The logarithmic divergence is clearly visible. Right: energy of the Coulomb gas (large deviation function up to a constant) obtained from the simulation compared to numerical integration of Eq.~(\ref{eq:energy}) using the densities (\ref{eq:rhoPh1},\ref{eq:rhoPh2}).}
  \label{fig:MC}
\end{figure}

The density obtained numerically matches perfectly our computation (see the left part of Fig.~\ref{fig:MC}). In particular, the logarithmic divergence is clearly visible.
The large deviation function $\Phi_\kappa(s)$ is also in perfect agreement with our results (Fig.~\ref{fig:MC}, right).
We stress that the comparison of the numerical and analytical results does not involve any adjustable parameter.

%%%%%%%%%%%%%%%%%%%%%%%%%%%%%%%%%%%%%%%%%%%%%%%%%%%%%%%%%%%%%%%%%%%%%%%%%%%%%%%%%%%%%%%%%% 
%%%%%%%%%%%%%%%%%%%%%%%%%%%%%%%%%%%%%%%%%%%%%%%%%%%%%%%%%%%%%%%%%%%%%%%%%%%%%%%%%%%%%%%%%% 

\section{Conclusion}
\label{sec:Conclusion}

\begin{figure}[!ht]
  \centering
  \includegraphics[width=0.6\textwidth]{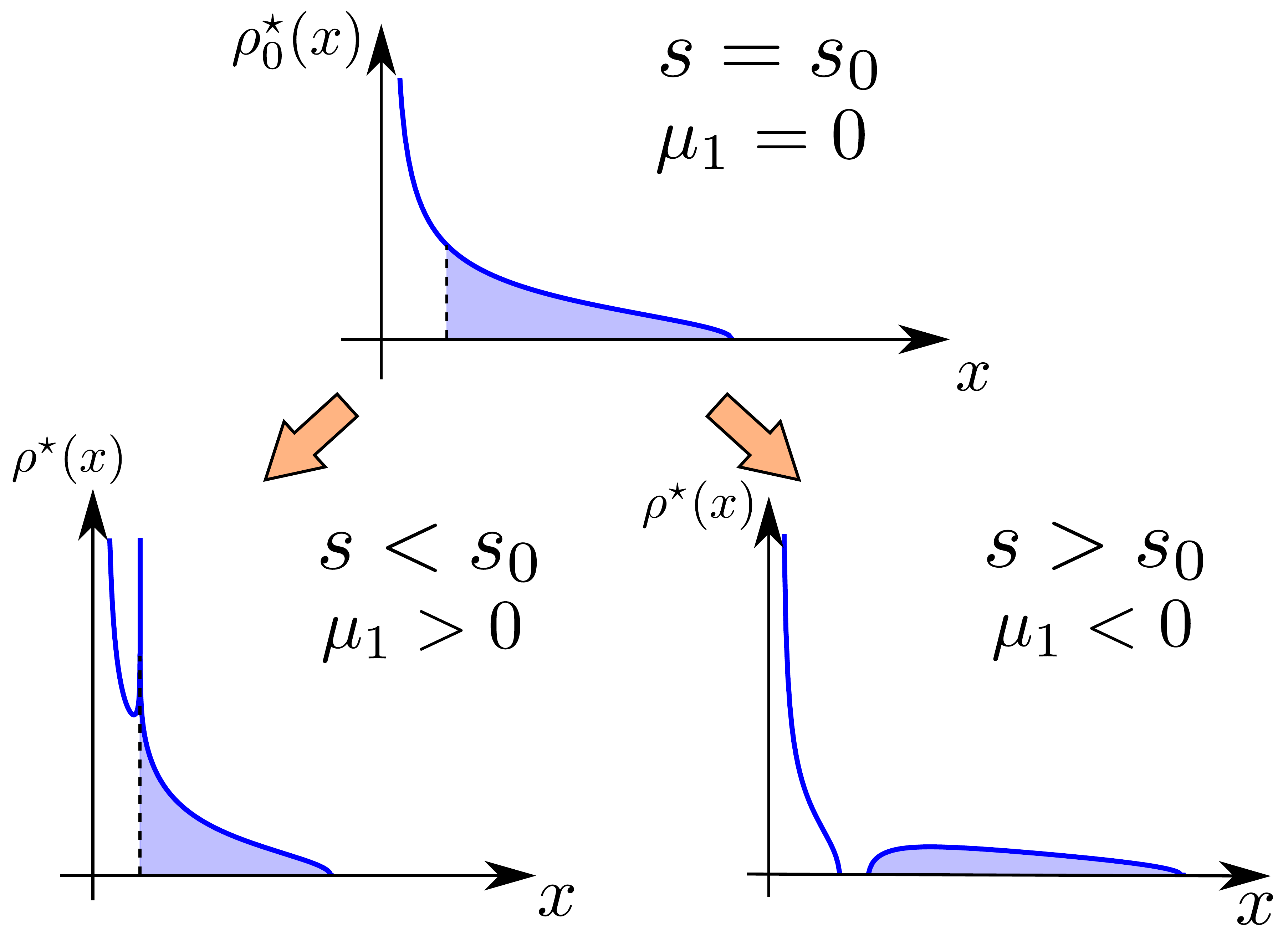}
  \caption{\it Sketch of the mechanism described in this paper. The optimal density $\rho^\star$ of the gas undergoes a phase transition driven by the constraint imposed by (\ref{eq:defTrLS}). $\rho_0^\star$ is the density obtained when the constraint is released. 
    % $s$ corresponds to the truncated linear statistics $\tilde{L}$ up to a rescaling with $N$. $\sz$ is the typical value $\tilde{L}_0$ up to the same rescaling, and $\lagS$ is the Lagrange multiplier introduced to handle the constraint.
  }
  \label{fig:Schema}
\end{figure}

In this paper we have studied an example of truncated linear statistics of \textit{top} eigenvalues $\tilde{L} = \sum_{n=1}^{\Ni} f(\lambda_n)$ within the Laguerre ensemble of random matrices.
In the large $N$ limit, by using the Coulomb gas technique, the problem has been recast as an optimization problem where we have searched for the most probable charge configuration consistent with the constraint that $s=N^{-\eta}\sum_{n=1}^{\Ni}f(\lambda_n)$ is fixed~; $\eta$ is an exponent ensuring that $s$ scales as $\mathcal{O}(N^0)$ when $N\to\infty$ (e.g. $\eta=3/2$ for $f(\lambda)=\sqrt{\lambda}$ in the Laguerre ensemble).
When the constraint is removed, the optimal charge density $\rhoz$ is the Mar\v{c}enko-Pastur density, which thus provides the typical (most probable) value of the truncated linear statistics:
\begin{equation}
  \sz(\kappa) = \int_{c_0} \rho_0^\star(x) f(x) \dd x,
  \quad 	\text{where} \quad
  \kappa =  \int_{c_0} \rho_0^\star(x) \dd x
  \:.
\end{equation}
These two equations define a line $s=\sz(\kappa)$ in the $(\kappa,s)$ plane, which was shown to correspond to a phase transition line of infinite order (essential singularity in the energy).
This line separates two phases characterized by different density profiles (Fig.~\ref{fig:PhDiag}):
on one side the density is supported on two disconnected intervals and on the other side the density has a compact support. In this second case, the density presents a logarithmic divergence \textit{inside} the bulk (the main mechanism is sketched in Fig.~\ref{fig:Schema}).
Although these new results were obtained within the Laguerre ensemble and for the specific choice of truncated linear statistics $\tilde{L}=\sum_{n=1}^{\Ni}\sqrt{\lambda_n}$ which has arisen from a model of Brownian interfaces, they are quite \textit{universal}. We have demonstrated in Appendix~\ref{app:sol} that the mechanism is independent of the choice of the function $f$, provided it is monotonous.
It is also independent of the matrix ensemble.
The case where the function $f$ is non monotonous is however still an open question (this is for example the case for the shot noise of chaotic cavity~\cite{VivMajBoh10}).

\begin{figure}[!ht]
  \centering
  \includegraphics[width=0.85\textwidth]{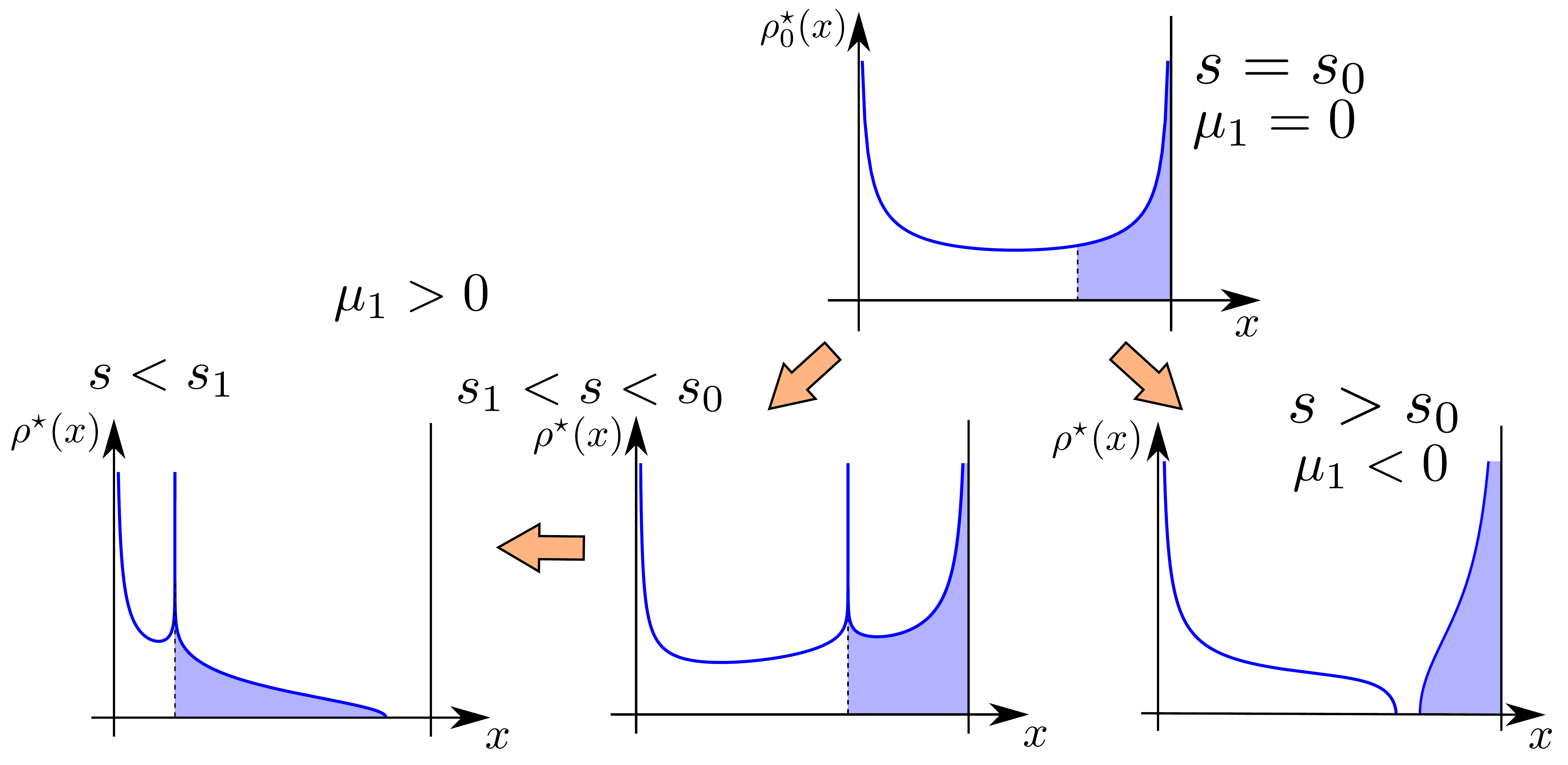}
  \caption{\it Sketch of the evolution of the density of eigenvalues in the Jacobi ensemble for different values of the linear statistics $s$ for $f(\lambda)=\lambda$. 
    For $s$ in the neighbourhood of the typical value $\sz(\kappa)$, the gas follows the same scenario as for the Laguerre ensebmle. There is however an additional third order transition at $s = s_1 < \sz$.}
  \label{fig:SchemaCcl}
\end{figure}

\begin{figure}[!ht]
  \centering
  \includegraphics[width=0.7\textwidth]{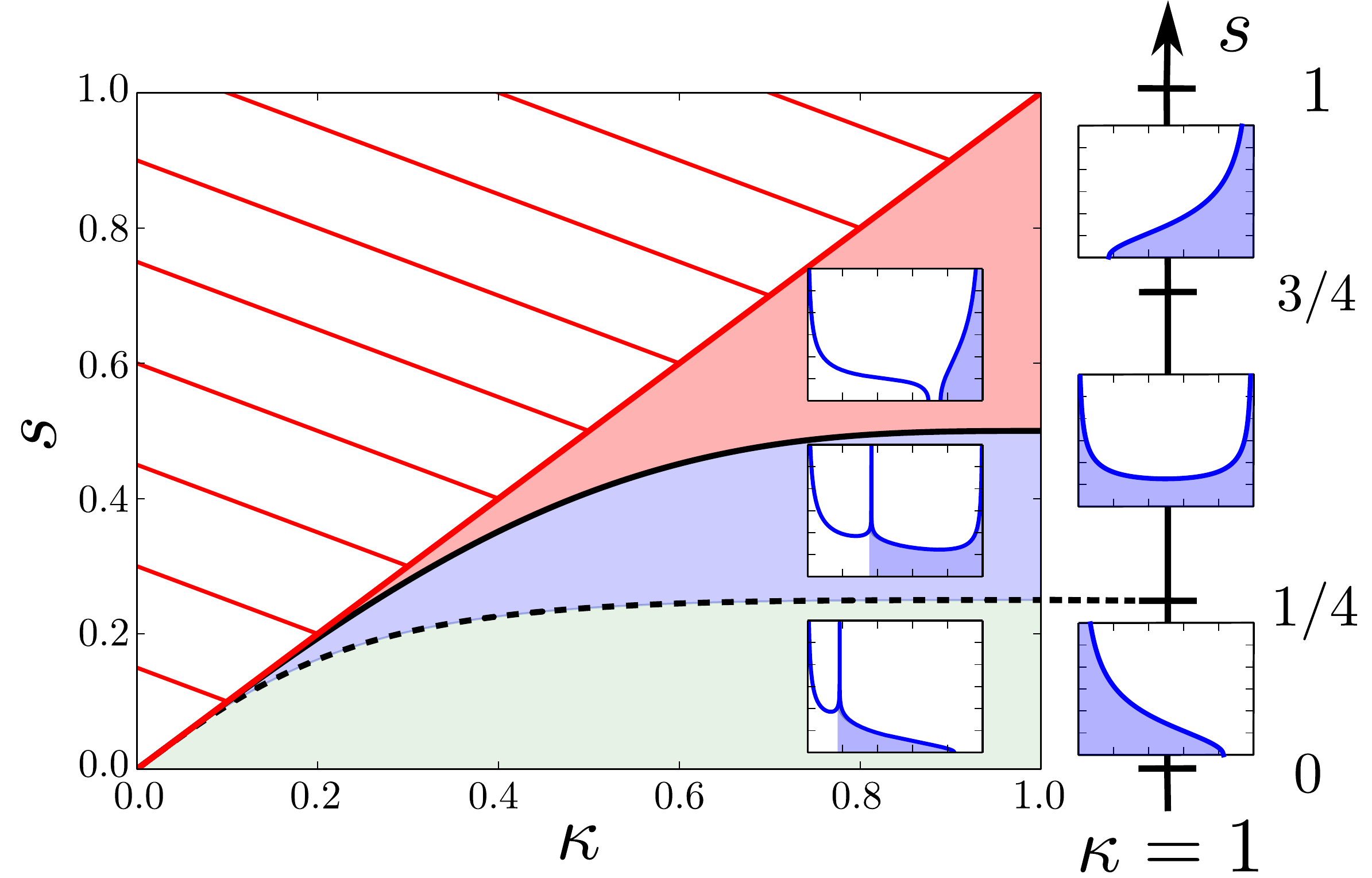}
  \caption{\it Phase diagram in the Jacobi ensemble for the linear statistics $f(x)=x$. 
    Since the eigenvalues are bounded, $s \in [0,\kappa]$ (the dashed region is the forbidden area).
    The black line corresponds to $s=\sz(\kappa)$, where the infinite order phase transition occurs. The dashed line, say $s=s_1(\kappa)$, corresponds to the phase transition from a hard edge to a soft edge, which is  third order.
    The right of the figure shows the phase diagram for $\kappa=1$ with the two third order phase transitions at $s=1/4$ and $s=3/4$.
    While the dashed line continues all the way to $\kappa=1$ (with $s_1(\kappa\to1)=1/4$), the solid line stops before $\kappa=1$ (with $\sz(\kappa\to1^-)=1/2$).
  }
  \label{fig:PhDiagJacobi}
\end{figure}

We stress that, depending on the function $f$ and the matrix ensemble, the whole picture may be richer than in the case considered in the paper.
We have also studied the truncated linear statistics $\tilde{L} = \sum_{n=1}^{\Ni}\lambda_n$ within the \textit{Jacobi} ensemble, corresponding to the joint probability density function
\begin{equation}
  P(\lambda_1, \ldots, \lambda_N) \propto \prod_{i<j} \abs{\lambda_i - \lambda_j}^\beta
  \prod_{n=1}^N \lambda_n^{\frac{\beta}{2}-1},
  \hspace{1cm}
  0 < \lambda_n < 1.
\end{equation}
Since the eigenvalues are bounded, $\tilde{L}$ now remains between $0$ and $\Ni$ and therefore $s=\tilde{L}/N \in [0, \kappa]$. 
The forbidden region $s > \kappa$ is dashed in Fig.~\ref{fig:PhDiagJacobi}.
The scenario introduced in the paper (Fig.~\ref{fig:Schema}) is only part of the full scenario as the constraint drives a second phase transition from a hard to a soft edge, as sketched in Fig.~\ref{fig:SchemaCcl} (this second transition is of third order~\cite{VivMajBoh10}). 
As a consequence the whole phase diagram is richer, as shown in Fig.~\ref{fig:PhDiagJacobi}.

We now make few remarks concerning the two limits $\kappa\to0$ and $\kappa\to1$.
In the case $\Ni=1$ with $f(\lambda)=\lambda$, the study of the truncated linear statistic distribution is exactly mapped onto the study of the largest eigenvalue, a problem which has been widely discussed in the literature~\cite{Joh00,Joh01,DeaMaj06,DeaMaj08,MajVer09,NadMaj09,NadMajVer11}.
However our Coulomb gas analysis in Section~\ref{sec:CoulombGas} cannot be used to consider the case $\kappa = 1/N$ as we have studied the thermodynamic limit $N\to\infty$ with $\kappa$ fixed and have only obtained the leading order contribution to the energy of the gas.

In Section~\ref{sec:DistrS}, the study of the Laguerre ensemble with $f(\lambda)=\sqrt{\lambda}$ has shown that the large deviation function for the truncated linear statistics ($\kappa<1$) continuously goes towards the one obtained in Ref.~\cite{NadMaj09} ($\kappa=1$).
We emphasize that this is \textit{not} a general feature.
For $0<\kappa<1$, the (infinite order) phase transition takes place at the typical value $s=\sz(\kappa)$.
On the other hand, for $\kappa=1$ (full linear statistics), the typical value $\sz(1)$ does not correspond in general to a phase boundary,~\footnote{
  The linear statistics $L=\sum_{n=1}^N \sqrt{\lambda_n}$ in the Laguerre ensemble is quite specific:
  the typical value corresponds to a phase transition and the density presents an unusual additional logarithmic behaviour at the hard edge, $\rho(x) \sim - \ln x/\sqrt{x}$ as $x \to 0$~\cite{NadMaj09}.
  hence the limit $\kappa\to1$ is in general also a singular limit.
  For example, for $L=\sum_{n=1}^N\lambda_n$ within the Jacobi ensemble, we clearly see in Fig.~\ref{fig:PhDiagJacobi} that the line of infinite order phase transition terminates at $\sz(1)=1/2$, which is below the third order phase transition occuring at $s=3/4$ when $\kappa=1$.
}

The singular nature of the two limits is clearly related to non commutation of the limits $\lim_{\kappa\to1}$ (or $\lim_{\kappa\to0}$) and $\lim_{N\to\infty}$.
In particular it would be interesting to study precisely the boundary of the phase diagram  for $\kappa\to1$ beyond the thermodynamic limit (the region of the phase diagram of Fig.~\ref{fig:PhDiagJacobi} where $\kappa\sim1$).
Such a general analysis would certainly be interesting as it is known that the Coulomb gas may present very diverse behaviours depending on the ensemble and the function $f$ (cf.~Table~\ref{tab:CGpt}).

\def\scale{0.1}
\def\scaleTwo{0.07}

\begin{table}[!ht]
  \begin{center}
    \begin{tabular}{lllcr}
      Quantity & Ensemble & order  & type & Ref. \\
      \hline 
      \hline 
      Wigner time delay  
      & Laguerre                   
      & second 
      & \diagram{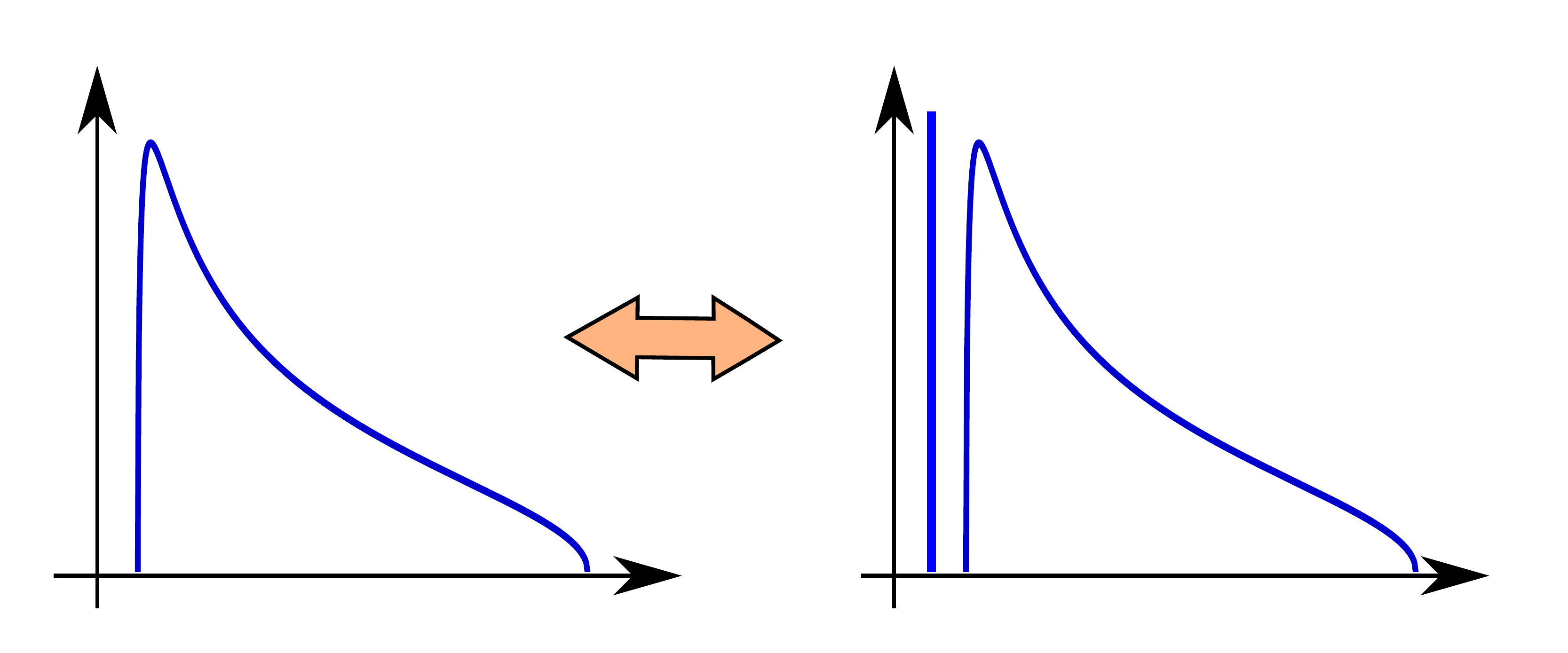}{\scale}{-0.75cm}  %single charge detached 
      & \cite{TexMaj13} 
      \\[-0.5cm]
      $\sum_i\lambda_i^{-1}$
      & 
      &
      & %from soft edge
      & 
      \\[0.1cm]
      \hline 
      Renyi entropy 
      & Laguerre
      & second 
      & \diagram{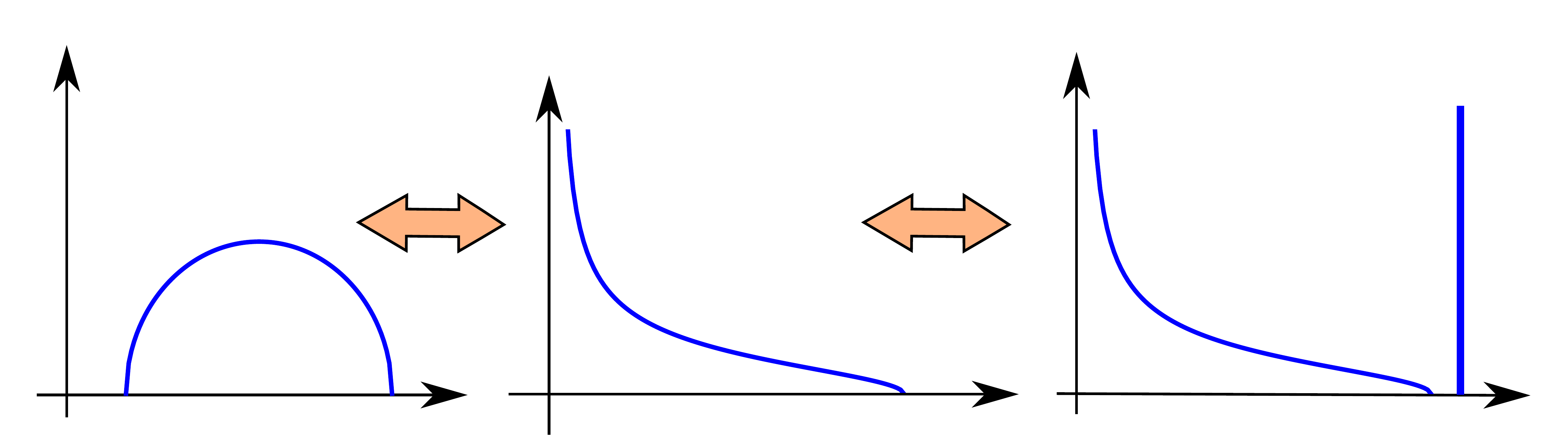}{\scale}{-0.75cm}  %single charge from soft edge
      & \cite{NadMajVer11}
      \\[-0.5cm]
      $\sum_i\lambda_i^{q}$
      & 
      & \& third
      &   
      & 
      \\[0.1cm]
      \hline 
      Conductance  
      & Jacobi
      & third 
      & \diagram{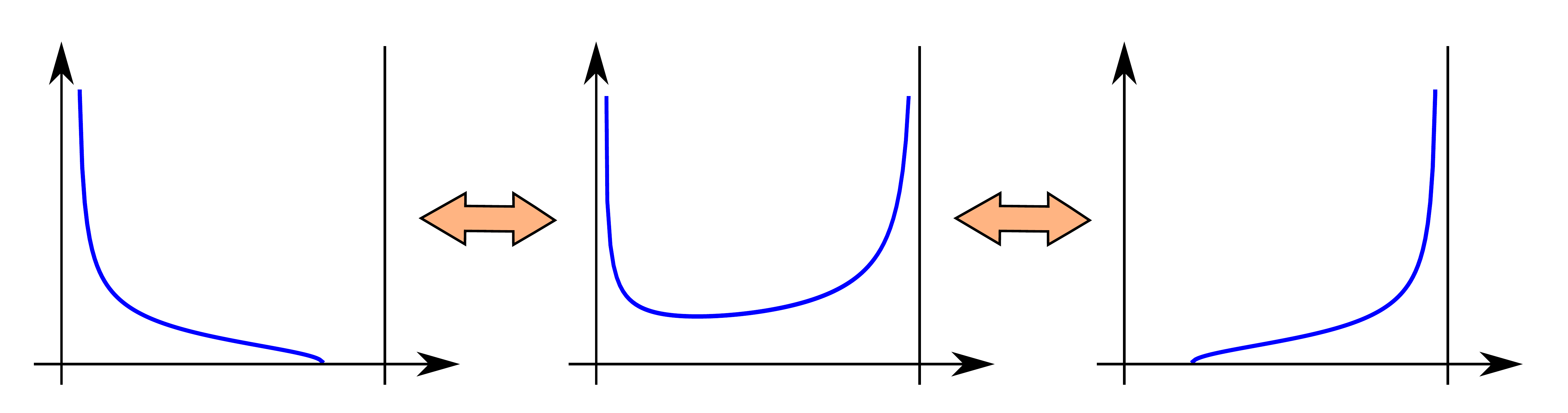}{\scale}{-0.75cm}  %hard edge to soft edge
      & \cite{VivMajBoh08,VivMajBoh10} 
      \\[-0.5cm]
      $\sum_i\lambda_i$
      & 
      & 
      & 
      & 
      \\[0.1cm]
      \hline 
      Shot noise, 
      & Jacobi 
      & third 
      & \diagram{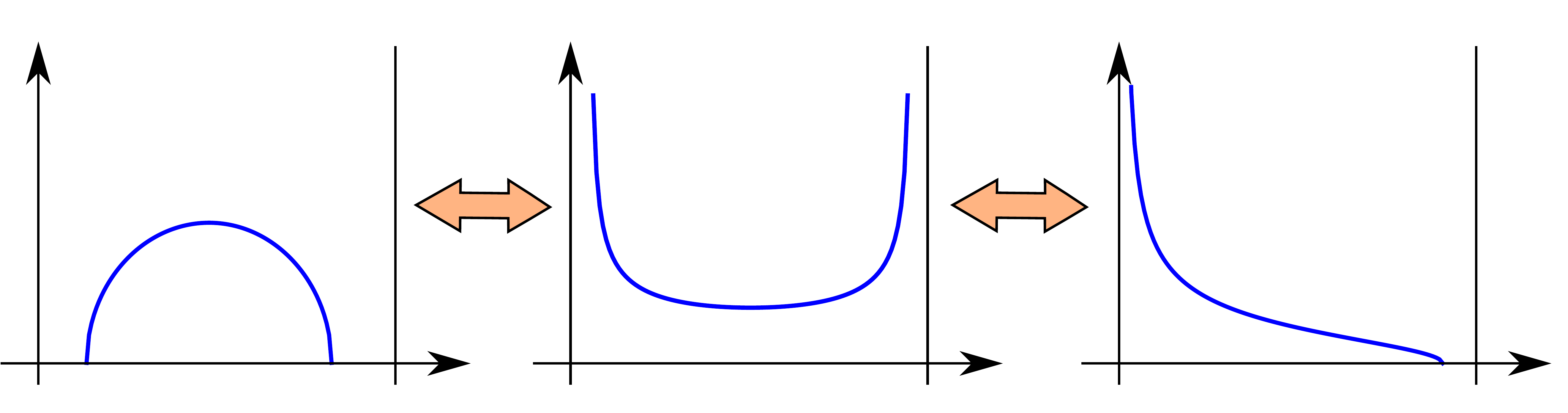}{\scale}{-0.75cm}  %hard edge to soft edge 
      & \cite{VivMajBoh10} 
      \\[-0.5cm]
      $\sum_n\lambda_n(1-\lambda_n)$ &
      &
      &
      & 
      \\[0.1cm]
      \hline
      Moments,  
      & Jacobi 
      & third 
      & \diagram{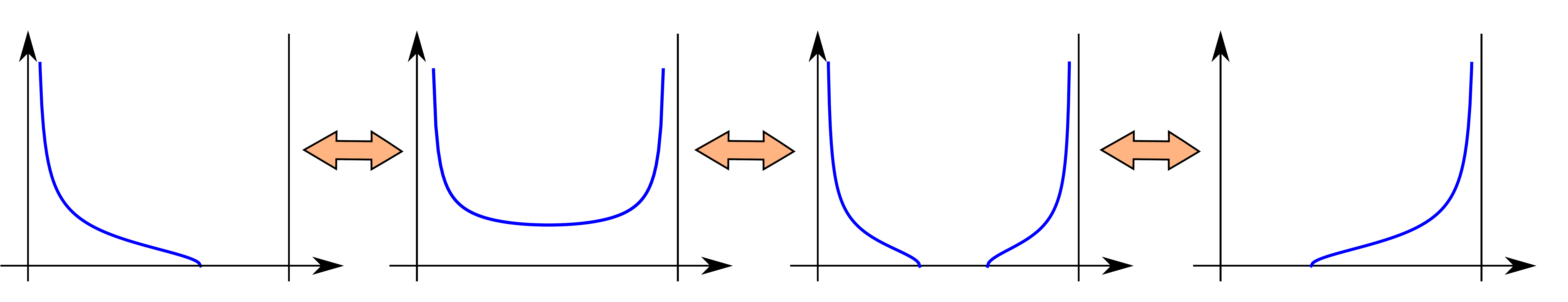}{\scaleTwo}{-0.6cm}  %gap opening with soft edges
      &  \cite{VivMajBoh10}
      \\[-0.5cm]
      $\sum_n\lambda_n^k$  &
      & 
      & 
      & 
      \\[0.1cm]
      \hline 
      NS conductance, 
      & Jacobi
      & third 
      & \diagram{TrMom}{\scaleTwo}{-0.75cm}  %same as higher moments
      &  \cite{DamMajTriViv11} 
      \\[-0.5cm]
      $\sum_n\lambda_n^2(2-\lambda_n)^{-2}$
      &
      &
      &
      & 
      \\[0.1cm]
      \hline 
      Largest eigenvalue  
      & Gaussian,            
      & third 
      & \diagram{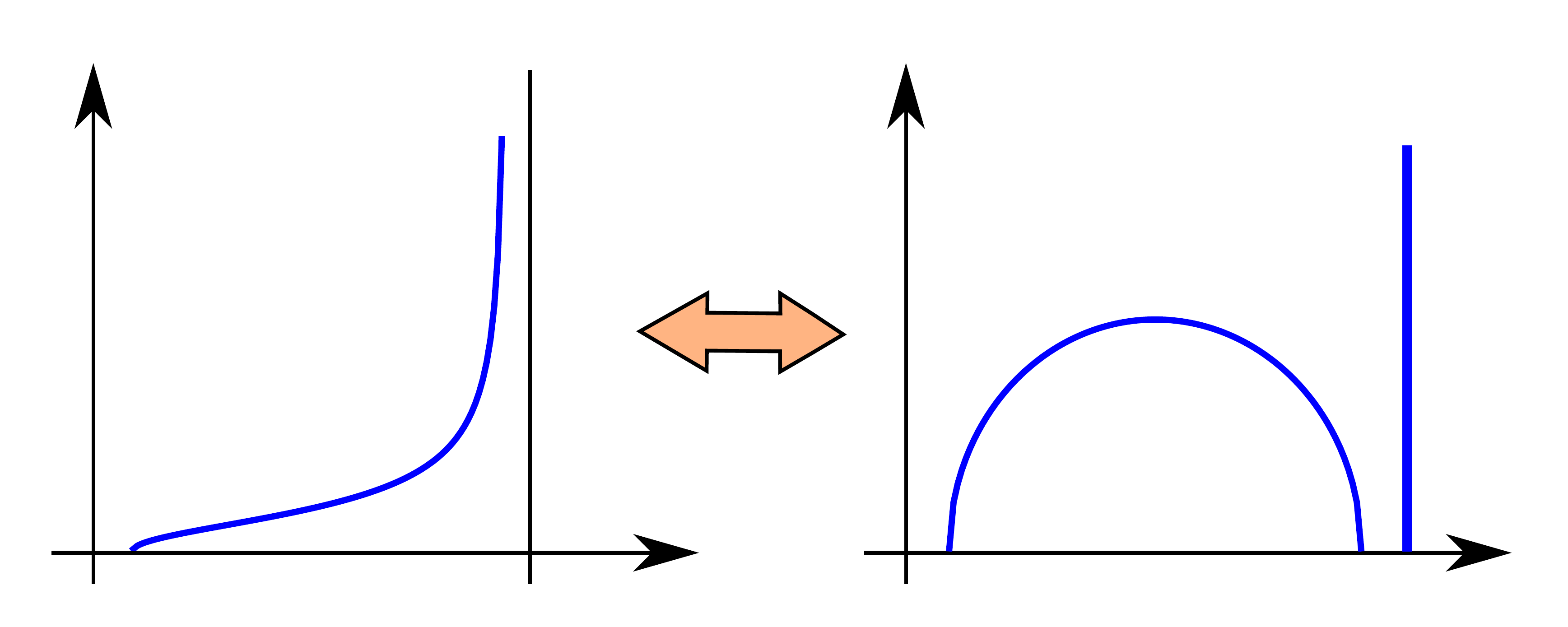}{\scale}{-0.75cm}  %hard edge to soft edge + charge
      & \cite{DeaMaj08} 
      \\[-0.5cm]
      $\lambda_1$
      & Laguerre       
      &
      &
      & \cite{MajVer09,MajSch14} 
      \\[0.1cm]
      \hline 
      Index  
      & Gaussian,  
      & 
      & \diagram{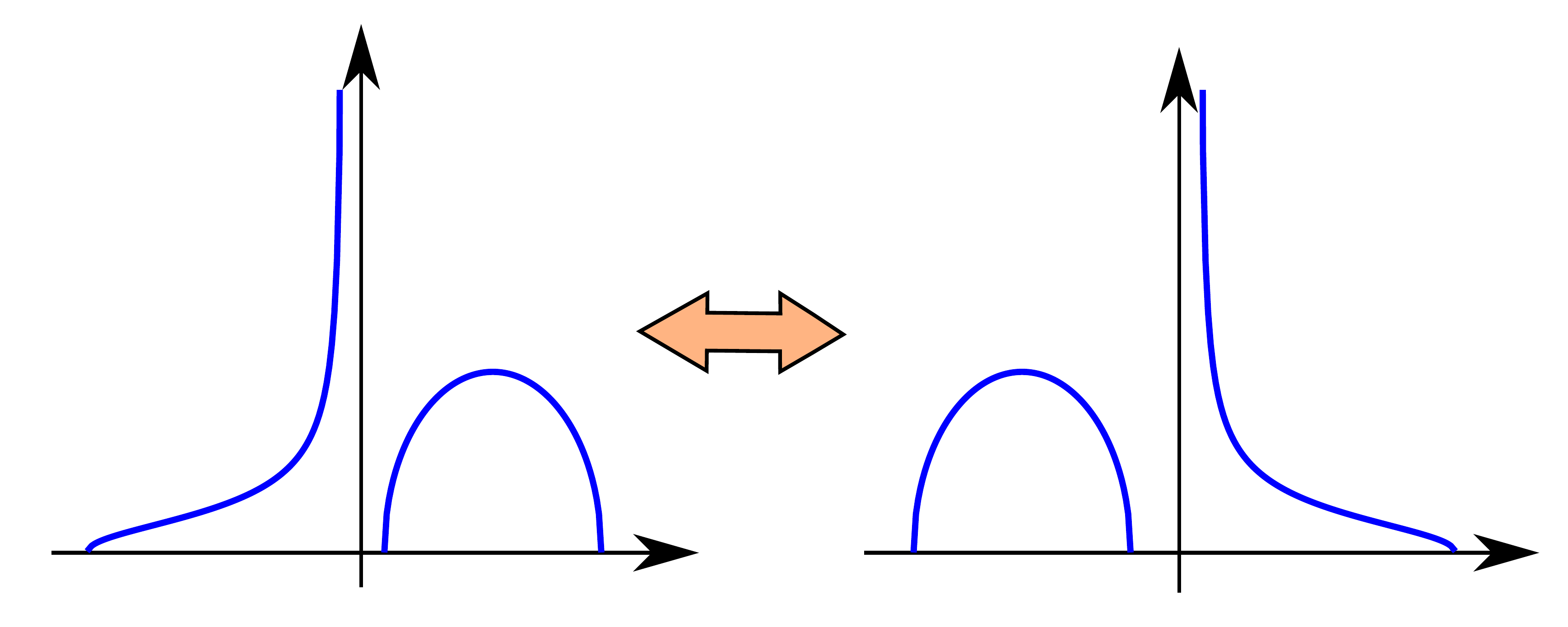}{\scale}{-1cm}  %soft/hard to hard/soft
      & \cite{MajNadScaViv09,MajNadScaViv11} 
      \\[-0.75cm]
      $\sum_n\Theta(\lambda_n)$
      & Laguerre
      & $(^*)$
      & 
      & \cite{MarMajSchViv14,MarMajSchViv14a}
      \\
      & Cauchy,...
      &
      &
      &
      \\[0.1cm]
      \hline 
      Index  (2D)
      & Complex
      & third
      & 
      &
      \cite{AlTouWain14}
      \\
      $\sum_n\Theta(\abs{\lambda_n}-r)$
      & Ginibre
      &
      &
      &
      \\[0.1cm]
      \hline
      Mean radius (2D)
      & Complex
      & fourth
      & 
      &
      \cite{CunMalMez15}
      \\
      $\sum_n|\lambda_n|$
      & Ginibre
      &
      &
      &
      \\[0.1cm]
      \hline  
      Center of mass 
      & Laguerre
      & infinite
      &  \diagram{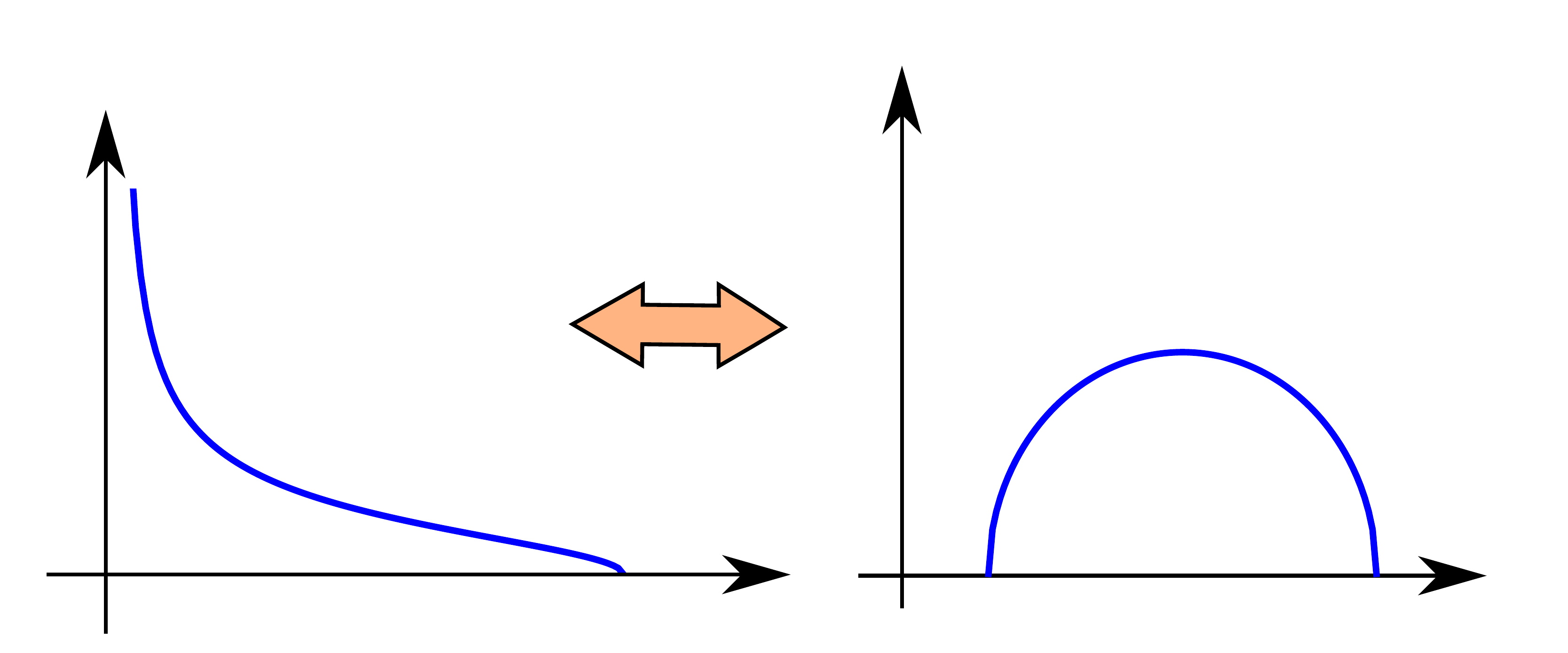}{\scale}{-0.75cm}
      & \cite{NadMaj09}
      \\[-0.5cm]
      of interf. $\sum_i\sqrt{\lambda_i}$
      & 
      &  $(^{**})$
      &   
      & 
      \\[0.1cm]
      \hline
      \textbf{Truncated} %linear statistics           
      & 
      & 
      & 
      & 
      \\[-0.3cm]
      $\sum_i^{\Ni}\sqrt{\lambda_i}$
      & Laguerre
      & infinite 
      &  \diagram{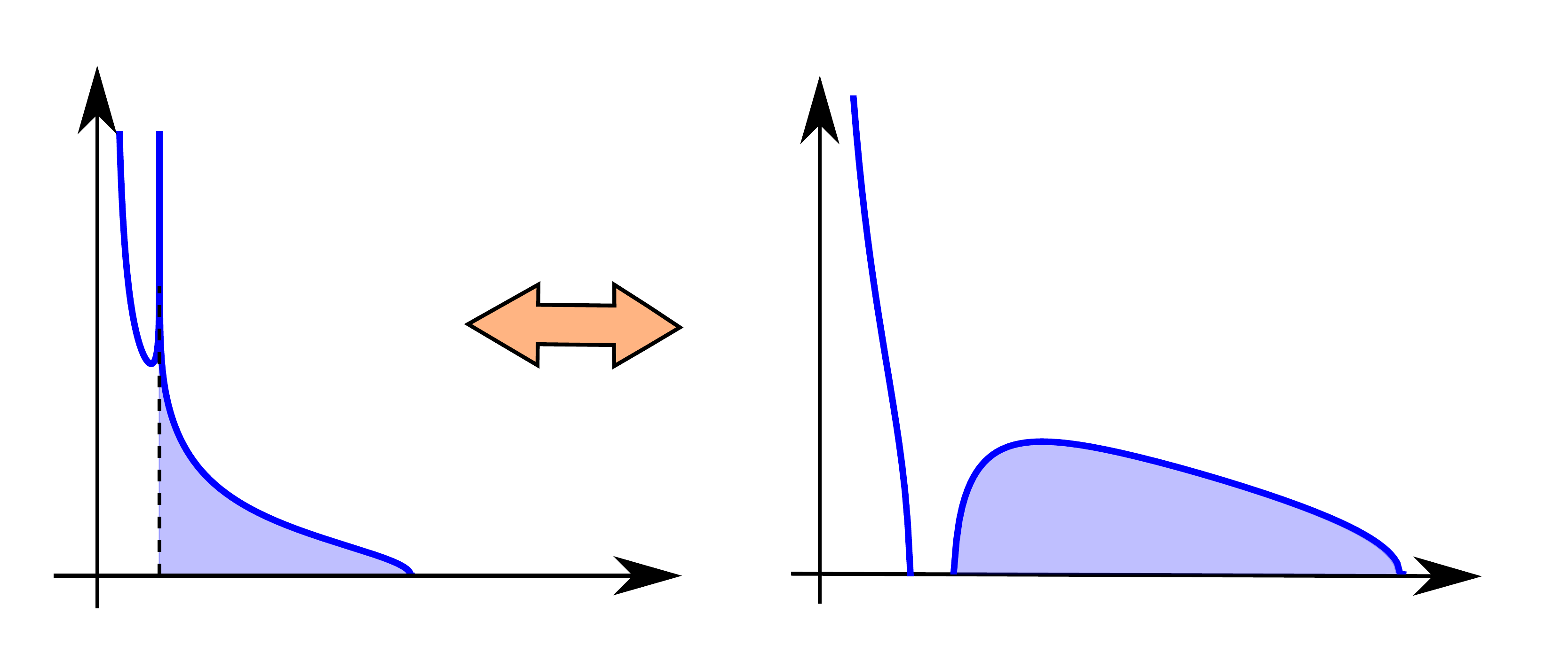}{\scale}{-0.75cm} 
      & this 
      \\[-0.5cm]
      $\sum_i^{\Ni}\lambda_i$
      & Jacobi
      & $\infty$ \& $3^\mathrm{rd}$
      & 
      & paper 
      \\[0.1cm]
      \hline 
      \hline
    \end{tabular}
  \end{center}
  \caption{\it List of different phase transitions observed in the Coulomb gas.
    $(^{*})$~: the energy has a logarithmic correction at the typical value.
    $(^{**})$~: the density presents a logarithmic correction at the hard edge 
    $\rho(x)\sim-\ln x/\sqrt{x}$ for $x\to0$. 
  }
  \label{tab:CGpt}
\end{table}

Much effort has been devoted to analyse the eigenvalue correlations in random matrices. Their universal character was underlined and has led to different types of correlations: sine kernel in the bulk of the density, Airy kernel in the vicinity of a soft edge and Bessel kernel in the vicinity of a hard edge~\cite{Meh04,For10}. 
The existence of a new type of behaviour in the density (logarithmic divergence inside the bulk) thus call for the study of the corresponding eigenvalue correlations.

Although the physical motivation for the analysis of truncated linear statistics presented here came from the model of Brownian interfaces, we expect that it also applies to other physical situations.
Cold atoms could be a possible field of application: 
it is known that the eigenvalues $\lambda_i$ of a matrix from the Gaussian Unitary ensemble correspond to the positions of 1D spinless free fermions in a harmonic well \cite{MarMajSchViv14a}. Hence, many physical quantities can be expressed as linear statistics of the $\lambda_i$'s. 
Any restriction of these quantities to a given number $\Ni$ of the rightmost fermions can therefore be treated as a truncated of the largest eigenvalues $\lambda_1, \ldots, \lambda_{\Ni}$, which might be of experimental interest with the recent progresses in measurement of the atom density~\cite{CheNicOkaGerRamBakWasLomZwi15,HalHudKelCotPeaBruKuh15,ParHubMazChiSetWooBlaGre15}.

\section*{Acknowledgements}

We are thankful to Gernot Akemann and Gr\'egory Schehr for interesting remarks.

%%%%%%%%%%%%%%%%%%%%%%%%%%%%%%%%%%%%%%%%%%%%%%%%%%%%%%%%%%%%%%%%%%%%%%%%%%%%%%%%%%%%%%%%%% 
%%%%%%%%%%%%%%%%%%%%%%%%%%%%%%%%%%%%%%%%%%%%%%%%%%%%%%%%%%%%%%%%%%%%%%%%%%%%%%%%%%%%%%%%%% 
%%%%%%%%%%%%%%%%%%%%%%%%%%%%%%%%%%%%%%%%%%%%%%%%%%%%%%%%%%%%%%%%%%%%%%%%%%%%%%%%%%%%%%%%%% 

\appendix

%%%%%%%%%%%%%%%%%%%%%%%%%%%%%%%%%%%%%%%%%%%%%%%%%%%%%%%%%%%%%%%%%%%%%%%%%%%%%%%%%%%%%%%%%% 
%%%%%%%%%%%%%%%%%%%%%%%%%%%%%%%%%%%%%%%%%%%%%%%%%%%%%%%%%%%%%%%%%%%%%%%%%%%%%%%%%%%%%%%%%% 

\section{Solution of the saddle point equation (general analysis within the Laguerre ensemble)}
\label{app:sol}

We study here a more general case than the one discussed in the body of the paper.
First, we consider the Laguerre ensemble of random matrix theory corresponding to the joint distribution of eigenvalues:
\begin{equation}
  P(x_1, \ldots, x_N) \propto
  \prod_{i<j} \abs{x_i - x_j}^\beta \prod_{n=1}^N x_n^{\beta N \nu/2} \EXP{-\beta N x_n/2},
  \hspace{1cm}
  x_n > 0
  \:,
\end{equation} 
where $\beta$ is the Dyson index (the eigenvalues have already been rescaled by a factor $N$ so they remain of order one in the limit $N \to \infty$). 
The exponent  $\nu \geq 0$ is now supposed of order $\mathcal{O}(N^0)$ (in the text, we considered $\nu=0$).
Note that the computation can be carried out in a similar way for other matrix ensembles.
Second, the function $f$ entering in the truncated linear statistics
\begin{equation}
  s = \frac{1}{N} \sum_{n=1}^{\Ni} f(x_n), \qquad
  x_1 > x_2 > \cdots > x_N
\end{equation}
is now an arbitrary monotonous function (the case $f(\lambda)=\sqrt{\lambda}$ was considered in the text.

Again, we study the limit $N \to \infty$ with $\kappa = \Ni/N$ fixed. Using the same notations as in Section~\ref{sec:CoulombGas}, the saddle point equation (\ref{eq:SteepestDescent}) becomes:
\begin{align}
  \label{eq:SDRho1b}
  2 \dashint \frac{\rhoR(y)}{x-y} \dd y + 2 \int \frac{\rhoL(y)}{x-y} \dd y
  &=  1 - \frac{\nu}{x} + \lagS f'(x) & \quad \text{for } x > \C
  \:, 
  \\
  \label{eq:SDRho2b}
  2 \int \frac{\rhoR(y)}{x-y} \dd y + 2 \dashint \frac{\rhoL(y)}{x-y} \dd y
  &= 1- \frac{\nu}{x} & \quad \text{for } x < \C
  \:,
\end{align}
Assuming the support of $\rhoL$ is $[a,b]$ and the support of $\rhoR$
is $[c,d]$, the second equation gives $\rhoL$ in terms of $\rhoR$
using Tricomi's theorem (\ref{eq:Tri1},\ref{eq:Tri2}):
\begin{equation}
  \label{eq:AppRhoL0}
  \rhoL(x) = \frac{1}{\pi \sqrt{(x-a)(b-x)}} 
  \left\lbrace
    A + \dashint_a^b \frac{\dd t}{\pi} \frac{\sqrt{(t-a)(b-t)}}{t-x}
    \left[ \frac{1}{2} \left( 1-\frac{\nu}{x} \right)
    - \int_c^d \frac{\rhoR(y)}{t-y} \dd y
    \right] 
  \right\rbrace
  \:.
\end{equation}
The constant $A$ is the normalisation of the density $\rhoL$,
\begin{equation}
  \label{eq:AppCteA}
  \int_a^b \rhoL(x) \dd x = A
  \:.
\end{equation}
The principal value integral can be computed by making use of the
following relations
\begin{equation}
  \label{eq:AppInt1}
  \dashint_a^b \frac{\dd t}{\pi} \frac{\sqrt{(t-a)(b-t)}}{t-x}
  = \frac{a+b}{2} - x
  \:,
\end{equation}
\begin{equation}
  \label{eq:AppInt2}
  \dashint_a^b \frac{\dd t}{\pi} \frac{\sqrt{(t-a)(b-t)}}{t-x}
  \frac{1}{t}
  = -1 + \frac{\sqrt{ab}}{x}
  \:,
\end{equation}
which hold for $0 \leq a < x < b$. The remaining term can be evaluated
by permuting the integrals:
\begin{equation}
  \dashint_a^b \frac{\dd t}{\pi} \frac{\sqrt{(t-a)(b-t)}}{t-x} 
  \int_c^d \frac{\rhoR(y)}{t-y} \dd y
  = \int_c^d \dd y \rhoR(y)
  \dashint_a^b \frac{\dd t}{\pi} \frac{\sqrt{(t-a)(b-t)}}{t-x}
  \frac{1}{t-y}
  \:.
\end{equation}
Using now
\begin{equation}
  \dashint_a^b \frac{\dd t}{\pi} \frac{\sqrt{(t-a)(b-t)}}{t-x}
  \frac{1}{t-y}
  = -1 + \frac{\sqrt{(y-a)(y-b)}}{y-x}
  \:,
  \quad
  \text{for}
  \quad
  y > b
  \:,
\end{equation}
we obtain:
\begin{equation}
  \label{eq:AppDblInt}
  \dashint_a^b \frac{\dd t}{\pi} \frac{\sqrt{(t-a)(b-t)}}{t-x} 
  \int_c^d \frac{\rhoR(y)}{t-y} \dd y
  = -\int_c^d \rhoR(y) \dd y
  + \int_c^d \rhoR(y) \frac{\sqrt{(y-a)(y-b)}}{y-x} \dd y
  \:.
\end{equation}
Using
Eqs.~(\ref{eq:AppCteA},\ref{eq:AppInt1},\ref{eq:AppInt2},\ref{eq:AppDblInt}),
we can rewrite Eq.~(\ref{eq:AppRhoL0}) as
\begin{align}
  \rhoL(x) = \frac{1}{\pi \sqrt{(x-a)(b-x)}} \bigg\lbrace \int_a^b
  \rhoL(x) \dd x +  \frac{1}{2} \left[ \frac{a+b}{2} - x - \nu \left(
      \frac{\sqrt{ab}}{x} - 1 \right) \right]
  \\
  +\int_c^d \rhoR(y) \dd y
  -\int_c^d \rhoR(y) \frac{\sqrt{(y-a)(y-b)}}{y-x} \dd y
  \bigg\rbrace
  \:.
\end{align}
Since $\int_a^b \rhoL(x) \dd x + \int_c^d \rhoR(x) \dd x = 1$, this
expression reduces to
\begin{align}
  \nonumber
  \rhoL(x) = \frac{1}{\pi \sqrt{(x-a)(b-x)}} \bigg\lbrace 
  1 
  + \frac{1}{2} &\left[ \frac{a+b}{2} - x - \nu \left( \frac{\sqrt{ab}}{x} - 1 \right) \right]
  \\
  & - \int_c^d \rhoR(y) \frac{\sqrt{(y-a)(y-b)}}{y-x} \dd y
  \bigg\rbrace
  \:.
  \label{eq:rho2A}
\end{align}
A divergence $\rho(x) \sim (b-x)^{-1/2}$ at the edge of a support is
usually caused by a hard wall, which is absent here at
$x=b$. Therefore we expect that the expression within the brackets
vanishes at $x=b$. Imposing this condition leads to the more compact
form:
\begin{equation}
  \rhoL(x) = \frac{1}{2\pi} \sqrt{\frac{b-x}{x-a}}
  \left\lbrace
    1 - \frac{\nu}{x} \sqrt{\frac{a}{b}} + 2 \int_c^d \sqrt{\frac{y-a}{y-b}} \frac{\rhoR(y)}{y-x} \dd y
  \right\rbrace
  \:.
\end{equation}
Plugging this expression into (\ref{eq:SDRho1b}) yields an equation on $\rhoR$ only:
\begin{equation}
  2\dashint \frac{\rhoR(t)}{x-t} \sqrt{\frac{t-a}{t-b}} \dd t = 1
  - \frac{\nu}{x} \sqrt{\frac{a}{b}} + \lagS f'(x) \sqrt{\frac{x-a}{x-b}}
  \:,
  \label{eq:EqRho2OnlyApp}
\end{equation}
which can be solved by applying Tricomi's theorem (\ref{eq:Tri1},\ref{eq:Tri2}) once more. The additional potential $\lagS f$ coming from the constraint either confines or pushed the charges away from the origin, hence the type of solution will depend on the sign of $\lagS f'$.
In the body of the paper we considered the case $f(x) = \sqrt{x}$. Therefore, it was only required to discuss the sign of $\lagS$. But here the sign of $f'$ is important to determine whether Phase~1 corresponds to $\lagS < 0$ (as before) or $\lagS > 0$, and similarly for Phase~2.

\subsection{Phase 1: $\lagS f' < 0$}

This corresponds to the situation in which the largest eigenvalues are pushed to the right, giving a density supported on two disjoint intervals (like Phase 1 in Fig.~\ref{fig:PhDiag}). 
Imposing that the densities vanish on the edges of their supports, we obtain:
\begin{align}
  \label{eq:Rho1Appendix}
  \rho_1(x) & = 
  \frac{1}{2\pi} \sqrt{\frac{(b-x)(d-x)}{(x-a)(c-x)}} \left\lbrace
    1 - \frac{\nu}{x} \sqrt{\frac{ac}{bd}} + \lagS
    \int_c^d \frac{\dd t}{\pi} \frac{f'(t)}{t-x} \sqrt{\frac{(t-a)(t-c)}{(t-b)(d-t)}}
  \right\rbrace
  \:,
  \\
  \label{eq:Rho2Appendix}
  \rho_2(x) & = 
  \frac{1}{2\pi} \sqrt{\frac{(x-b)(d-x)}{(x-a)(x-c)}} \left\lbrace
    1 - \frac{\nu}{x} \sqrt{\frac{ac}{bd}} + \lagS
    \dashint_c^d \frac{\dd t}{\pi} \frac{f'(t)}{t-x} \sqrt{\frac{(t-a)(t-c)}{(t-b)(d-t)}}
  \right\rbrace
  \:,
\end{align}
along with the conditions:
\begin{equation}
  \hspace{-1cm}
  1 - \nu \sqrt{\frac{a}{bcd}} + \lagS \int_c^d \frac{\dd t}{\pi} \sqrt{\frac{t-a}{(t-b)(t-c)(d-t)}} f'(t) =0
  \label{eq:A1cdt1}
  \:,
\end{equation}
deduced by imposing $\rhoR(c) = 0$, 
and 
\begin{equation}
  1 + \frac{\nu}{2} + \frac{a+c-b-d}{4} - \frac{\nu}{2} \sqrt{\frac{ac}{bd}}
  - \frac{\lagS}{2} \int_c^d \frac{\dd t}{\pi} \sqrt{\frac{(t-a)(t-c)}{(t-b)(d-t)}} f'(t) = 0
  \label{eq:A1cdt2}
  \:,
\end{equation}
obtained in the derivation
of~(\ref{eq:Rho1Appendix},\ref{eq:Rho2Appendix}) when imposing that
the solution of~(\ref{eq:EqRho2OnlyApp}) vanishes at $x=d$. One
should now distinguish two cases:
\begin{itemize}
\item $\nu > 0$: in this case $a>0$ and the condition $\rhoL(a)=0$ is explicitely
  \begin{equation}
    1 - \nu \sqrt{\frac{c}{a b d}} 
    + \lagS \int_c^d \frac{\dd t}{\pi} \sqrt{\frac{t-c}{(t-a)(t-b)(d-t)}} f'(t) = 0
    \:.
    \label{eq:A1cdt3}
  \end{equation}
\item $\nu=0$: the boundary of the support is the origin, $a=0$.
\end{itemize}
The parameters $c$ and $\lagS$ are fixed by the constraints
\begin{equation}
  \int \rhoR(x) \dd x = \kappa 
  \quad \mbox{ and } \quad
  \int \rhoR(x) f(x) \dd x = s
  \:.
  \label{eq:ConstraintsA}
\end{equation}

\subsection{Phase 2: $\lagS f' > 0$}

In this case, the eigenvalues are pushed towards the origin, and the two densities merge: $b=c$  (like Phase 2 in Fig.~\ref{fig:PhDiag}). Similarly, we obtain:
\begin{equation}
  \rho_1(x) = 
  \frac{1}{2\pi} \sqrt{\frac{d-x}{x-a}}  \left\lbrace
    1 - \frac{\nu}{x} \sqrt{\frac{a}{d}}
    + \lagS
    \int_c^d \frac{\dd t}{\pi} \sqrt{\frac{t-a}{d-t}} \frac{f'(t)}{t-x}
  \right\rbrace
  \:,
\end{equation}
\begin{equation}
  \rho_2(x) = 
  \frac{1}{2\pi} \sqrt{\frac{d-x}{x-a}}  \left\lbrace
    1 - \frac{\nu}{x} \sqrt{\frac{a}{d}}
    + \lagS
    \dashint_c^d \frac{\dd t}{\pi} \sqrt{\frac{t-a}{d-t}} \frac{f'(t)}{t-x}
  \right\rbrace
  \:.
\end{equation}
To obtain these expressions, we have imposed that the general solution
of~(\ref{eq:EqRho2OnlyApp}) vanishes for $x=d$, which gives the
condition
\begin{equation}
  1 + \frac{\nu}{2} + \frac{a-d}{4} - \frac{\nu}{2} \sqrt{\frac{a}{d}} = 
  \frac{\lagS}{2} \int_c^d \frac{\dd t}{\pi} \sqrt{\frac{t-a}{d-t}} f'(t)
  \:.
\end{equation}
\begin{itemize}
\item $\nu > 0$: in this case $a>0$ and we impose $\rhoL(a)=0$,
  \begin{equation}
    1 - \frac{\nu}{\sqrt{a d}} + \lagS \int_c^d \frac{\dd t}{\pi} \frac{f'(t)}{\sqrt{(t-a)(d-t)}}=0
    \:.
  \end{equation}
\item $\nu=0$: the support's boundary is $a=0$.
\end{itemize}
Again, the parameters $\lagS$ and $c$ are fixed by the constraints (\ref{eq:ConstraintsA}). In addition, when $x \to c^-$,
\begin{equation}
  \int_c^d \frac{\dd t}{\pi} \sqrt{\frac{t-a}{d-t}} \frac{f'(t)}{t-x} 
  \simeq - \frac{f'(c)}{\pi} \sqrt{\frac{c-a}{d-c}} \ln \abs{x-c}
  \:.
\end{equation}
This equation is also valid if one considers the same integral with $x>c$, provided the integral is a Cauchy principal value integral. 
This shows that the density of eigenvalues exhibits a logarithmic divergence at $x=c$:
\begin{equation}
  \rhoS(x) \simeq - \frac{\lagS}{2} f'(c) \ln \abs{x-c}.
\end{equation}
This analysis demonstrates that the presence of the logarithmic singularity is not specific to the model studied in the body of the paper, but a direct consequence of the restriction of the linear statistics $s$ to the largest eigenvalues.

\subsection{Phase transition of infinite order}
\label{sec:AppPhTr}

We again interpret the transition between these two types of densities as a phase transition for the Coulomb gas. The transition occurs when $\lagS=0$, and the corresponding density of eigenvalues is given by the Mar\v{c}enko-Pastur law:
\begin{equation}
  \rhoS_0(x) = \frac{\sqrt{(x-x_-)(x_+-x)}}{2\pi x}
  \quad\mbox{where }
  x_\pm = 2+\nu \pm 2 \sqrt{1+\nu}
  \:.
\end{equation}
The value of the parameter $c$ is then given by
\begin{equation}
  \kappa = \int_c^{x_+} \rhoS_0(x) \dd x
  \:,
\end{equation}
and the corresponding value of the truncated linear statistics is
\begin{equation}
  s = \sz(\kappa) = \int_c^{x_+} \rhoS_0(x) f(x) \dd x
  \:.
\end{equation}

Let us study the case $\nu > 0$. In the first phase, the limit $\lagS f'\to 0^-$ corresponds to the case where the two supports merge ($ b \to c $). In this limit, Eqs.~(\ref{eq:A1cdt1},\ref{eq:A1cdt2},\ref{eq:A1cdt3},\ref{eq:ConstraintsA}) reduce to:
\begin{align} 
  &\lagS \simeq \frac{\alpha}{\ln (c-b)}
  \qquad\mbox{ where  }
  \alpha = 
  \frac{\pi}{f'(c)} \sqrt{\frac{d-c}{c-a}} \left( 1 - \frac{\nu}{c}\sqrt{\frac{a}{d}} \right)
  \:,
  \label{eq:AlphaApp}
  \\
  &1 + \frac{\nu}{2} + \frac{a-d}{4} - \frac{\nu}{2} \sqrt{\frac{a}{d}} = 
  \frac{\lagS}{2} \int_c^d \frac{\dd t}{\pi} \sqrt{\frac{t-a}{d-t}} f'(t)
  + \mathcal{O}(\EXP{\alpha/\lagS}),
  \\
  &1 - \frac{\nu}{\sqrt{a d}} + \lagS \int_c^d \frac{\dd t}{\pi} \frac{f'(t)}{\sqrt{(t-a)(d-t)}}= \mathcal{O}(\EXP{\alpha/\lagS})
  \:,
  \\
  \nonumber
  &\kappa 
  = \int_c^d \frac{\dd x}{2\pi} \sqrt{\frac{d-x}{x-a}} \left( 1 - \frac{\nu}{x} \sqrt{\frac{a}{d}} \right)
  \\
  &\hspace{0.5cm}
  + \lagS \int \frac{\dd x}{2\pi} \sqrt{\frac{d-x}{x-a}} \dashint_c^d \frac{\dd t}{\pi} \sqrt{\frac{t-a}{d-t}} \frac{f'(t)}{t-x}
  + \mathcal{O}(\EXP{\alpha/\lagS}/\lagS)
  \:,
  \\
  \nonumber
  &s = \int_c^d \frac{f(x) \dd x}{2\pi} \sqrt{\frac{d-x}{x-a}} \left( 1 - \frac{\nu}{x} \sqrt{\frac{a}{d}} \right)\\
  &\hspace{0.5cm}
  + \lagS \int \frac{f(x) \dd x}{2\pi} \sqrt{\frac{d-x}{x-a}} \dashint_c^d \frac{\dd t}{\pi} \sqrt{\frac{t-a}{d-t}} \frac{f'(t)}{t-x}
  + 
  \mathcal{O}(\EXP{\alpha/\lagS}/\lagS)
  \label{eq:SMu1App2}
  \:.
\end{align}
Performing a similar expansion for the second phase, i.e. for $\lagS f'\to 0^+$, we obtain the same equations without the $\mathcal{O}(\EXP{\alpha/\lagS})$ and $\mathcal{O}(\EXP{\alpha/\lagS}/\lagS)$ terms.
But these corrections give no contribution to any power series at $\lagS = 0$, hence all the derivatives of the energy (related to $\lagS$ via \eqref{eq:thermoId}) are equal on both sides of the transition. There is a weak non analyticity (an essential singularity) at $\lagS = 0$, corresponding to $s = \sz(\kappa)$. Using the standard terminology of phase transition, we can say that the Coulomb gas undergoes a phase transition of infinite order at $s = \sz(\kappa)$.

Coming back to the case $f(x) = \sqrt{x}$ with $\nu=0$
discussed in the main body of the paper, the discussion above proves
the result of Section~\ref{sec:PhTr}. Indeed,
Eq.~(\ref{eq:AlphaApp}) reduces to
\begin{equation}
  \alpha = 2 \pi \sqrt{4-c_0}
  \:.
\end{equation}
To recover Eq.~(\ref{eq:EssSingPhi}), we need to rewrite the terms
$\mathcal{O}(\EXP{\alpha/\mu_1})$ and
$\mathcal{O}(\EXP{\alpha/\mu_1}/\mu_1)$ in terms of $\epsilon =
s-s_0$. This can be done using Eq.~(\ref{eq:SMu1App2}), which
reduces to
\begin{equation}
  \mu_1 = - \frac{2\pi^2}{4-c_0 + c_0 \ln (c_0/4)} \epsilon
  + \mathcal{O}(\epsilon^2) 
  + \mathcal{O}(\EXP{\alpha/\mu_1}/\mu_1)
  \:,
\end{equation}
where we have kept the subleading
$\mathcal{O}(\EXP{\alpha/\mu_1}/\mu_1)$ as it is the one that
differs between Phases 1 and 2. From this relation, we deduce  
\begin{equation}
  \frac{\alpha}{\mu_1} \simeq -\frac{\sqrt{4-c_0}}{\pi \epsilon}
  \left( 4-c_0 + c_0 \ln \frac{c_0}{4} \right)
  = - \frac{\gamma}{\epsilon}
  \:,
  \quad
  \text{for}
  \quad
  \mu_1 \to 0
  \:,
\end{equation}
where $\gamma$ is the constant given in Eq.~(\ref{eq:ExprGamma}). We
can thus obtain the energy (or equivalently the large deviations
function $\Phi_\kappa$) by using the thermodynamic
identity~(\ref{eq:thermoId}). Since the only difference between the
two phases is the $\mathcal{O}(\EXP{\alpha/\mu_1}/\mu_1) =
\mathcal{O}(\EXP{-\gamma/\epsilon}/\epsilon)$, we have, for
$\epsilon>0$:
\begin{equation}
  \frac{\dd}{\dd \epsilon} 
  \left[
    \Phi_\kappa(s_0+\epsilon) - \Phi_\kappa(s_0-\epsilon)
  \right]
  = \mathcal{O}(\EXP{-\gamma/\epsilon}/\epsilon)
  \:.
\end{equation}
Integrating this last relation yields~(\ref{eq:EssSingPhi}).

%%%%%%%%%%%%%%%%%%%%%%%%%%%%%%%%%%%%%%%%%%%%%%%%%%%%%%%%%%%%%%%%%%%%%%%%%%%%%%%%%%%%%%%%%% 
%%%%%%%%%%%%%%%%%%%%%%%%%%%%%%%%%%%%%%%%%%%%%%%%%%%%%%%%%%%%%%%%%%%%%%%%%%%%%%%%%%%%%%%%%% 

% \bibliographystyle{phreport}
% \bibliography{Biblio-Aurelien}
% \end{document}

\end{document}